\documentclass[a4paper,fleqn,usenatbib]{mnras}

\usepackage[T1]{fontenc}
\usepackage{ae,aecompl}

\usepackage{graphicx}	% Including figure files
\usepackage{amsmath}	% Advanced maths commands
\usepackage{amssymb}	% Extra maths symbols

\usepackage{epsfig,graphics,bm}
\usepackage{subfigure}
\usepackage{txfonts}
\usepackage{tabulary} 
\usepackage[utf8]{inputenc}

\title[The stability of the outer Oort Cloud]{The stability in the most external region of the Oort Cloud: The evolution of the ejected comets.}

\author[Correa-Otto {\rm{\&}} Calandra]{
J. A. Correa-Otto,\thanks{E-mail: jorgecorreaotto@conicet.gov.ar}
M. F. Calandra,\thanks{E-mail: mfcalandra@conicet.gov.ar}
\\
Grupo de Ciencias Planetarias, Dpto. de Geof{\'i}sica y Astronom{\'i}a, Facultad de Ciencias Exactas, F{\'i}sicas y Naturales, Universidad Nacional de San Juan - CONICET, \\
 Av. J. I. de la Roza 590 oeste, J5402DCS Rivadavia, San Juan, Argentina\\
}

\date{Accepted XXX. Received YYY; in original form ZZZ}

\pubyear{2019}

\begin{document}
\label{firstpage}
\pagerange{\pageref{firstpage}--\pageref{lastpage}}
\maketitle

% Abstract of the paper
\begin{abstract}
In this paper, we present a study about the dynamical effects of the Galaxy on the external region of the Oort Cloud. The aims of this paper are: i) to determine an outer limit for the Oort Cloud; and ii) to analyse the dynamical behaviour of the most external objects of the Cloud and how they are ejected from the Solar System. This is undertaken by following the temporal evolution of massless test particles in the Galactic environment of the solar neighbourhood. Here we show that the effect of the perturbations from the Galactic tide in the particles is similar to that find for the evolution of wide binary stars population.
Moreover, in the Oort Cloud we found a dynamical structure around 10$^5$ au conformed by objects unbound of the Sun. This structure allows us to define a transition region of stability and an outer boundary for the Oort Cloud, and it is also in agreement with previous results about the disruption of wide binary stars. 
% Moreover, we found an important number of minor bodies with heliocentric separation of $\sim$ 10$^5$ au  which conforms an unbound Oort Cloud around the Sun.  Finally, we have estimated an outer limit for the Oort Cloud:  particles with initial semimajor axes larger than $\sim$ 6.5 $\times$ 10$^4$ au have a possibility of survive for 5 Gyr which is less than 5 \%.
%\textcolor{red}
\end{abstract}

% Select between one and six entries from the list of approved keywords.
% Don't make up new ones.

%%%%%%%%%%%%%%%%%%%%%%%%%%%%%%%%%%%%%%%%%%%%%%%%%%

%%%%%%%%%%%%%%%%% BODY OF PAPER %%%%%%%%%%%%%%%%%%

\section{Introduction}\label{intro}

In $1950$ Oort (1950) proposed the existence of a spherical cloud of icy objects around the Sun. This structure called now Oort Cloud would be the reservoir of observed long-period comets. The Cloud is probably the remnant of planetary formation, a process that places the comets at such large heliocentric distances through a combination of planetary and external perturbations.

This large and low density structure has an estimated mass between 2 and 40 $m_\oplus$ (Francis 2005), and it is believe to be formed by $10^{10}$ to $10^{12}$ icy bodies larger than 2.3 km (Weissman 1996; Brasser \& Morbidelli 2013). These small objects are isotropically distributed and orbit the Solar System outside the planetary region with perihelia larger than 32 au and semi-major axes ($a$) between 3 $\times$ 10$^{3}$ and 10$^{5}$ au (Dones et al. 2015). Moreover, the density profile of the Cloud is roughly a power law proportional to $r^{-3.5}$, where $r$ is the heliocentric distance (Duncan et al. 1987; Fouchard et al. 2017). However, the perihelia of Oort Cloud objects are usually driven into the planetary region by the effects of external perturbations such as passing stars (Rickman 1976; Rickman et al. 2008; Fernandez 1980; Fouchard et al. 2011a,b), the tidal field of the Milky Way (Byl 1983; Heisler \& Tremaine 1986), and encounters with giant molecular clouds(Hut \& Tremaine 1985; Jakubik \& Neslusan 2009).

There are several theoretical studies about the origin of the Oort Cloud, which consider different scenarios for its formation process. Dones et al. (2015) reviewed this topic in detail. There still is a great debate how the Oort Cloud formed, but it is believed that it was an interplay between planetary scattering and external influences, and this process took place during the first 0.5 Gyr of the Solar System evolution. On the other hand, several authors modeled different environments for the primordial Solar System. Brasser \& Morbidelli (2013) took into account the actual position of the Sun in the Galaxy for the external influence and assumed that the formation of the Oort Cloud starts with the giant planet migration according to the Nice model (Tsiganis et al. 2005). Other authors considered a migration of the Sun in the Galaxy, so that the external environment changes frequently with consequences for the evolution of the Cloud (Brasser et al. 2010; Kaib et al. 2011), or analyzed the possibility of an early formation of the Oort Cloud when the Sun was still in its birth cluster (Fernandez \& Brunini 2000; Brasser et al. 2006; Kaib \& Quinn 2008).

Another interesting topic related to the Oort Cloud is its shape and limits. This large structure can be divided into an outer and inner Oort Cloud formed by objects with semi-major axes larger or smaller than 2 $\times$ 10$^{4}$ au respectively, where the distinction between the two regions is supported by considerations of the evolution of cometary orbits (Hills 1981; Duncan et al. 1987), and the  boundary between them has been defined by the minimum semimajor axis a comet must have to be sufficiently perturbed by Galactic  tides  or  stellar  encounters  to  enter  the inner  Solar System. 

The Inner Oort Cloud has been more studied in the past (see, Dones et al. 2004, 2015), while the most external part of the Oort Cloud is more complicated to analyze. However, the discovery of the first interstellar minor body 1I/2017~U1 (`Oumuamua, Bacci et al. 2017; Meech et al. 2017a,b)  has induced new research about the subsequent evolution of the icy bodies when they left the Solar System (e.g., Hanse et al. 2018). In any case, there are several questions without a clear answer about the dynamical evolution of the outer Oort Cloud and its exterior limit.

For stability studies in the most external regions of the Solar System at semimajor axes larger than 5 $\times$ 10$^4$ au, it would be incorrect to assume that the object disappears instantaneously when its orbit becomes unbound. This is because the restricted two-body potential (i.e., Sun-comet) is no longer valid at so large separations, and the three-dimensional Galactic tidal field becomes significant. Then, in this region, the dynamics of small bodies are dominated by the Galactic potential (Heisler \& Tremaine 1986; Jiang \& Tremaine 2010; Correa-Otto et al. 2017), and the existence of an unbound Oort Cloud is possible. This cloud would be formed by icy bodies unbound from the Solar System that eventually will be ejected.

The aim of this paper is to study the stability of the most external objects of the Oort Cloud, which are under the effects of the gravitational potential of the Galaxy and passing stars, in order to improve our understanding about the outer dynamical limit of our Solar System. In section \ref{system} we define the initial configuration for the Cloud. In Sect. \ref{model} we describe the numerical methods employed for our dynamical study. In Sect. \ref{result} we present our results and we analyze the stability of the objects. Finally, discussions and conclusions close the paper in Sect. \ref{conclu}.

%*******************************************************
%*******************************************************
\section{Initial conditions for the objects of the Oort cloud}\label{system}
%*******************************************************
%*******************************************************

To analyze the stochastic effect of stellar passages we consider three different synthetic Oort Clouds formed by $10^6$ massless particles, we call them Sample 1, 2 and 3. We also considered that the bodies orbit the Sun in a coordinate system $(x, y, z)$, where the reference plane is the Galactic plane. The positive $z-axis$ is perpendicular to the Galactic plane and points towards the South Galactic Pole. Besides, in this heliocentric coordinate system, the reference line is the positive $x-axis$, which points radially outwards from the Galactic center, and therefore, the positive $y-axis$ points in the direction of the Galactic rotation. As the Sun orbits around the Galactic center, the particles are in a rotating not-inertial coordinate system.

The inclination ($I$) of the orbital plane of each massless particle is defined concerning the Galactic plane, and the longitude of the ascending node ($\Omega$) is defined from the positive $x-axis$. The eccentricity ($e$) define the shape of the orbit and the angular position of the perihelion is given by the argument of perihelion ($\omega$). Finally, we identify by $a$ the semimajor axis of the orbit, and the mean anomaly ($M$) indicates the position of the particle in its orbit.

The initial shape of the Cloud was generated following the standard model of a thermalized Oort Cloud described in Rickman et al. (2008) and Hanse et al. (2018), which assumes a spherically symmetric and isotropic distribution with initial radial density profile $\propto$ $r_0^{-3.5}$, where $r_0$ is the initial distance to the Sun (Duncan et al. 1987; Dybczynski 2002; Fouchard et al. 2011b, 2014; Feng \& Bailer-Jones 2014). Therefore, for the initial angular orbital elements and the initial cosine of inclination we assume uniform distributions, and for the initial distributions of the semimajor axis ($a_0$) and the eccentricity ($e_0$) we considered a probability density proportional to $a_0^{-1.5}$ and $e_0$, respectively. The lower and upper limits for distribution of $a_0$ are 3 $\times$ 10$^{3}$ au and  10$^{5}$ au, respectively, while in order to keep the initial orbit of the particles outside the planetary region we set an upper limit for $e_{0}$: $e_{max}=1-35\, (a_0)^{-1}$.

In our simulations, we do not include the planets because the outer part of the Cloud is dynamically governed by the external perturbations, so in a first approximation we can ignore any planetary effect. Therefore, the test particles are removed from the simulation when their heliocentric distances became smaller than 35 au because we can not predict the posterior evolution of such objects accurately. Moreover, to study the outer region of the Cloud, we have not defined an outer threshold, so that we can follow the fictitious minor bodies after they become unbound.

%*******************************************************
%*******************************************************
\section{Numerical test for the stability of the Oort Cloud}\label{model}  
%*******************************************************
%*******************************************************

To determine the stability of the particles of the three samples, we perform several numerical simulations searching for objects that can survive during a period similar to the estimated age of the Solar System ($\sim$ 5 Gyr). To solve the exact equations of motion, we used a Bulirsch–Stoer integrator with an adaptive step size, and an error tolerance of 10$^{-13}$.

In our numerical simulations we included the disturbing effects of the Galactic tidal field and passing stars. However, we do not took into account other perturbative effects such as encounters with molecular clouds or changes in the density of the Galactic environment (Brasser et al. 2010; Kaib et al. 2011). Then, it is worth to mention that the stability of the Cloud could be affected by these more powerful perturbations.

 \begin{figure}
 \centering
\subfigure{\includegraphics[width=0.99\columnwidth]{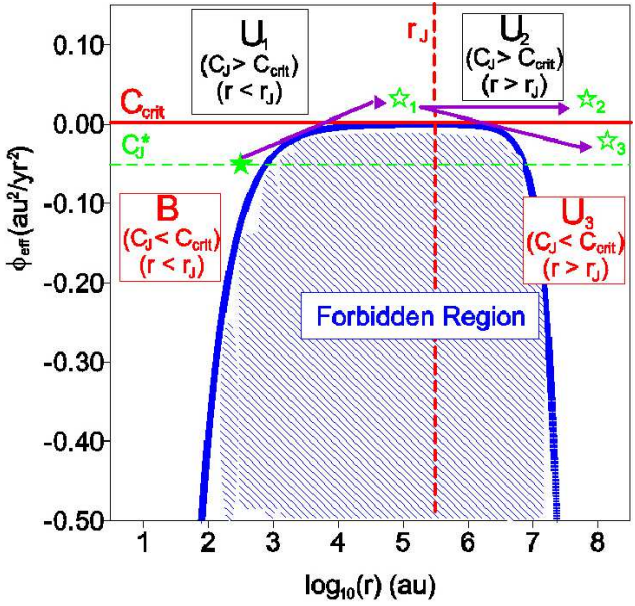}}
%\begin{center}
%\epsfig{figure=esquema3.eps,width=0.99\columnwidth} 
\caption{Effective potential as function of the heliocentric distance $r$. Blue line shows the potential, which has a maximum at $r \sim$ 2.8 $\times$ 10$^{5}$ au (dashed red line). The value of the maximum is $\sim$ -2.125 $\times$ 10$^{-4}$ au$^2$ yr$^{-2}$, and it is indicated by a continuous red line. Objects with $C_J > C_{crit}$ are able to move by all the distances, instead that particles with $C_J < C_{crit}$ have a region where the quadratic velocity is negative and there is a forbidden region (striped area). We schematically divide the picture in four regions: $B$, $U_1$, $U_2$ and $U_3$, where only that particles in the region $B$ are bound to the Sun.}
\label{fig1}
%\end{center}
\end{figure}

For the coordinate system described in Sect. \ref{system} the tidal field can be analytically modeled by the Hill’s approximation (Heggie 2001; Binney \& Tremaine 2008), assumimg a symmetric Galactic potential on the plane $z = 0$ (Jiang \& Tremaine 2010). Let $x$, $y$, $z$ denote the components of the heliocentric position of the massless particles, and $\dot{x}$, $\dot{y}$, $\dot{z}$ the components of the velocity. Then, from Jiang \& Tremaine (2010) the equations of motion for the particles are:

\begin{equation}
\begin{array}{lcl}
\ddot{x} &=& - \dfrac{\mathcal{G} m_0 x}{r^{3}} + 2 \Omega_G  \dot{y} + 4 \Omega_G A_G x \,  \rm{,}\\ 
\\
\ddot{y} &=& - \dfrac{\mathcal{G} m_0 y}{r^{3}} - 2 \Omega_G  \dot{x}  \,  \rm{,}\\ 
\\
\ddot{z} &=& - \dfrac{\mathcal{G} m_0 z}{r^{3}} - \nu_G^2 z  \,  \rm{,}\\ 
\label{eq1}
\end{array}
\end{equation}
\noindent where $\mathcal{G}$ is the gravitational constant and  $r=\sqrt{x^2+y^2+z^2}$. The effects of the Galactic tide are represented by the terms involving $\Omega_G$, $A_G$ and $\nu_G$ which are the angular speed of the Galaxy, the Oort constant, and the frequency for small oscillations in $z$, respectively. For the distance of the Sun to the Galactic centre ($\sim$ 8 kpc) their approximate values are (Jiang \& Tremaine 2010): $A_G^{-1} \sim 2 \, \Omega_G^{-1} \sim 4.8 \, \nu_G^{-1} \sim $ 6.61 $\times$ 10$^7$ yr.

Equations (\ref{eq1}) have a stationary solution (Jiang \& Tremaine 2010):
\begin{equation}
\begin{array}{l}
\ddot{x}\,=\, \ddot{y}\,=\, \ddot{z} \,=\, 0 \rm{,} \\ 
\\ 
\ddot{x}\,=\, \dot{y}\,=\, \dot{z} \,=\, 0 \rm{,} \\ 
\\ 
y \, = \, z \, = 0 \rm{,} \\ 
\\ 
x\, = \pm r_J = \pm \left[  \dfrac{\mathcal{G} m_0 }{4 A_G \Omega_G} \right]  ^{1/3} \,  \rm{,} \label{eq3}
\end{array}
\end{equation} 

\noindent where $r_J \sim$ 2.8 $\times$ 10$^{5}$ au is the Jacobi or tidal radius of the test particles. Moreover, the system Sun-particle admits one integral of motion, the Jacobi constant: 
\begin{equation}
\begin{array}{lcl}
C_J & = & \dfrac{1}{2} (\dot{x}^2 + \dot{y}^2 + \dot{z}^2) -  \dfrac{\mathcal{G} m_0 }{r} - 2 A_G \Omega_G x^2 + \dfrac{\nu_G^2}{2} z^2  \,\,  \rm{,} \\ 
 & & \\ 
    & = & \dfrac{1}{2} v^2  +  \phi_K + \phi_G   \,\, \rm{,} \\ 
     & & \\ 
    & = & \dfrac{1}{2} v^2 +  \phi_{eff} \,\,  \rm{,} \label{eq4}
\end{array}
\end{equation} 

\noindent where $v^2 = \dot{x}^2 + \dot{y}^2 + \dot{z}^2$ is the velocity, and

\begin{equation}
\begin{array}{l}
 \phi_K = - \dfrac{\mathcal{G} m_0 }{r}  \\ 
  \phi_G = - 2 A_G \Omega_G x^2 + \dfrac{\nu_G^2}{2} z^2  \rm{,} \label{eq5}
 \end{array}
\end{equation} 
are the potentials of the restricted two-body Keplerian problem and that of the Galactic tide, while $\phi_{eff} = \phi_K + \phi_G $ is the effective potential. Therefore, $\phi_{eff}$ has a maximum in the stationary solution (eq. \ref{eq3}), which is called the critical Jacobi constant:
\begin{equation}
C_{crit} = -3 \left[ \dfrac{A_G \Omega_G (\mathcal{G} m_0)^2}{2} \right]  ^{1/3} \rm{,} \label{eq6}
\end{equation} 

\noindent which for the objects of the Oort Cloud is $C_{crit} \sim$ -2.125 $\times$ 10$^{-4}$ au$^2$ yr$^{-2}$. For the Keplerian two-body problem the particles will be unbound of the Sun at $r \sim \infty$ (or $e\sim1$), however, if we consider the effective potential the separation can occur at a smaller distance. Moreover, for high values of $C_J$ close to the Sun (i.e., $r<$  0.5 $r_J$) the Keplerian potential can define a bound eccentric orbit, when in fact it corresponds to an unbound particle.

Figure \ref{fig1} shows the effective potential $\phi_{eff}$ as function of the heliocentric distance.  The particles with $C_J > C_{crit}$ are able to reach any distance close to or far from the Sun, while for those particles with $C_J < C_{crit}$ there is a range of distances where the motion is forbidden because the kinetic energy is negative, which is indicated in the figure with a striped area in blue. Fig. \ref{fig1} is also a helpful tool to understand the dynamical limit for the particles orbiting the Sun. We can define four regions: i) region $B$, for particles with $C_J < C_{crit}$ and $r < r_J$, ii) region $U_1$, for particles with $C_J > C_{crit}$ and $r < r_J$, iii) region $U_2$, for particles with $C_J > C_{crit}$ and $r > r_J$, and iv) region $U_3$, for particles with $C_J < C_{crit}$ and $r > r_J$. Only the objects in region $B$  will remain bound to the Sun in absence of external perturbations. The objects in regions $U_2$ and $U_3$ have been ejected of the Solar System, while in region  $U_1$ the particles are unbound of the Sun but still not ejected.

For the test particles in region $B$ and $U_1$ with distance $r \leq 10^5$ au the motion can be approximated by a disturbed two-body problem (Heisler \& Tremaine 1986; Correa-Otto et al. 2017). For the other two regions ($U_2$ and $U_3$) the particles are unbound and far from the Sun, so the Keplerian potential can be ignored, and the motion equations (\ref{eq1}) can be reduced to: 

\begin{equation}
\begin{array}{lcl}
\ddot{x} &=&  2 \Omega_G  \dot{y} + 4 \Omega_G A_G x \,  \rm{,}\\ 
\\
\ddot{y} &=&  - 2 \Omega_G  \dot{x}  \,  \rm{,}\\ 
\\
\ddot{z} &=& - \nu_G^2 z  \,  \rm{,}\\ 
\label{eq7}
\end{array}
\end{equation}

\noindent whose general solution is:

\begin{equation}
\begin{array}{lcl}
x(t) &=& x_0 + x_1 \cos(\kappa_G t + x_2) \,  \rm{,}\\ 
\\
y(t) &=&  y_0 - 2 A_G x_0 t - \dfrac{2 \Omega_G}{\kappa_G} x_1 \sin(\kappa_G t + x_2)  \,  \rm{,}\\ 
\\
z(t) &=& z_0 \cos(\nu_G t + z_1)  \,  \rm{,}\\ 
\label{eq8}
\end{array}
\end{equation}

\noindent where $x_0$, $x_1$, $x_2$, $y_0$, $z_0$ and $z_1$ are arbitrary constants, $\nu_G$ is the frequency for small oscillations in $z$, and $\kappa_G = 2 \sqrt{\Omega_G (\Omega_G - A_G)}$  is the epicyclic frequency in the $x$-$y$ plane. Therefore, the result is a uniform secular motion along the $y$-axis in the direction of the Galactic rotation.

On the other hand, the particles can be separated from the Sun by the cumulative effect of stellar passages, which correspond to a stochastic perturbation. Although the high relative velocities between the stars in the solar neighbourhood allows to include this perturbation using a model of impulse approximation (Rickman 1976), we prefer to solve each encounter with a second star, $m_1$, by a direct numerical integration of a restricted three body problem with the additional perturbation of the Galactic potential.

For the total time of integration ($T=$ 5 Gyr) and the maximum impact parameter, defined by the extension of our Clouds ($q_M \sim$ 1 pc), we found $N = 4 \times 10^4$ stellar passages with background stars (see, Brunini \& Fernández 1996; Jiang \& Tremaine 2010; Correa-Otto \& Gil-Hutton 2017), which are randomly distributed along the simulation. The frequency of encounters considered in our work is 8 Myr$^{-1}$, which is smaller than that considered in other recent works (e.g., Vokrouhlicky et al. 2019). So, our frequency of encounter can be considered as an approximation because it  is an underestimation of a more realistic encounter rate. 
% that provides an order of magnitude of the real effect, in any case the results could be scaled to any other frequency taking a proportion.

To generate the stellar passages we followed the scheme developed by Rickman et al. (2008), which is described in detail in Section 2 of that work. To select the mass $m_1$ of the passing star we use the mass-luminosity function in the solar neighbourhood (Reid et al. 2002; Ninkovic \& Trajkovska 2006) and the initial relative velocity for the encounter is taken from the velocity dispersion of nearby stars available in the Hipparcos data (Garcia-Sanchez et al. 2001), which is a function of the stellar masses. Then, we have the information needed to generate the random sequence of stellar passages. However, the scheme of Rickman et al. (2008) to generate stellar passages assume a isotropic distribution of the stellar velocities in the solar neighbourhood, which is not true due to the solar apex motion. This assumption in the scheme to generate encounters represent an approximation in our simulations.

It is worth to mention that the recent release of the Gaia DR2 data (Bailer-Jones et al. 2018) has updated the rate of encounters with different stellar types/classes in the solar neighbourhood. However, we are in the same situation of Vokrouhlicky et al. (2019), because these data had been published without any debugging when we started our simulations, which took several months due to  our limited computing capacity. Therefore, we will discuss the implications of this approximation for our results in Sect. \ref{conclu}.

Finally, to be able to develop a statistical study of the stochastic effect of stellar passages, we would like to perform several sequences of stellar passages. However, once again, we are restricted by our computer capabilities  because our simulations demand very much CPU time. So, we have randomly generated only three different sequences of stellar passages from the same encounter distribution. The sequence \textit{i} is apply to the Sample \textit{i}, with \textit{i}=1, 2 and 3, so that we have performed three numerical experiments.

%following the work of \cite{f11b}

\begin{table}

\centering
\begin{tabular}{c | c  c | c c  }
%\hline
  Sample   &   Final   & percentage  & Final & percentage   \\
           & state     &     \%      & region  & \%   \\
          \hline 
           \hline 
              &  & & & \\
  &    bound & 53 & B & 53   \\
   & & & & \\
  &     &  & U$_1$ & 1   \\
1 &    unbound & 34 & U$_2$ & 20   \\
  &     &  & U$_3$ & 13   \\
     & & & & \\
  &    eliminated & 13 & - & -   \\
     & & & & \\
  \hline
     & & & & \\
    &    bound & 70 & B & 70   \\
       & & & & \\
  &     &  & U$_1$ & 1   \\
2 &    unbound & 16 & U$_2$ & 14   \\
  &     &  & U$_3$ & 1   \\
     & & & & \\
  &    eliminated & 14 & - & -   \\
     & & & & \\
    \hline
       & & & & \\
    &    bound & 68 & B & 68   \\
       & & & & \\
  &     &  & U$_1$ & 1   \\
3 &    unbound & 18 & U$_2$ & 11.5   \\
  &     &  & U$_3$ & 5.5   \\
     & & & & \\
  &    eliminated & 14 & - & -   \\
     & & & & \\
%\hline
\end{tabular}
\caption{Final results of our simulations for the three samples. Second and third columns separate the final result of each Sample between the \textit{bound}, \textit{unbound} and \textit{eliminated} particles. Moreover, fourth and fifth columns separate the subsample of \textit{unbound} particles according to the schematic description of Fig. \ref{fig1}. }
\label{table1}
\end{table}

\section{Results}\label{result}

The Galactic effects inject particles to the inner Solar System, which become comets, but also eject particles from the Solar System, which is the topic of interest in this work.  Previous works considered that a particle escapes of the Solar System when the heliocentric distance is larger than $r_J$ (see, Sect. 3, Jiang \& Tremaine 2010; Fouchard et al. 2011b, 2017). However, in this work we follow the evolution of the particles until the end of each integration allowing them to reach distances larger than $r_J$ to study the stability of the most external particles of the Oort Cloud.

Table \ref{table1} shows the final results of our simulations for the three samples. We can separate our results in three groups, i) particles that enter the planetary region, crossing the threshold of 35 au and are \textit{eliminated} from the simulation, ii)  the objects that remains \textit{bound} to the Sun after 5 Gyr, and iii)  the particles ejected from the Solar System, which are \textit{unbound} and remain evolving in the Galaxy. The percentage of these three groups are in column 3 of Table \ref{table1}, and in fourth and fifth columns the results for the \textit{unbound} particles (third group) are detailed according to the schematic description of Fig. \ref{fig1}.

We found that at the end of the integration time the percentage of particles that reach the limit of 35 au is $\sim$ 14 \% for the three samples. Therefore, our results indicate that the eliminated particles by crossing the threshold of 35 au are not very strongly affected by the sequence of stellar passages. Instead, the particles ejected from the Solar System and the objects that remain \textit{bound} to the Sun after 5 Gyr, shown different percentages for the three samples, which indicate a dependence with the sequence of stellar passages. For example, the percentage of \textit{unbound} particles in Sample 1 is the double of that in the other two. 

 \begin{figure}
\centering
\subfigure{\includegraphics[width=0.99\columnwidth]{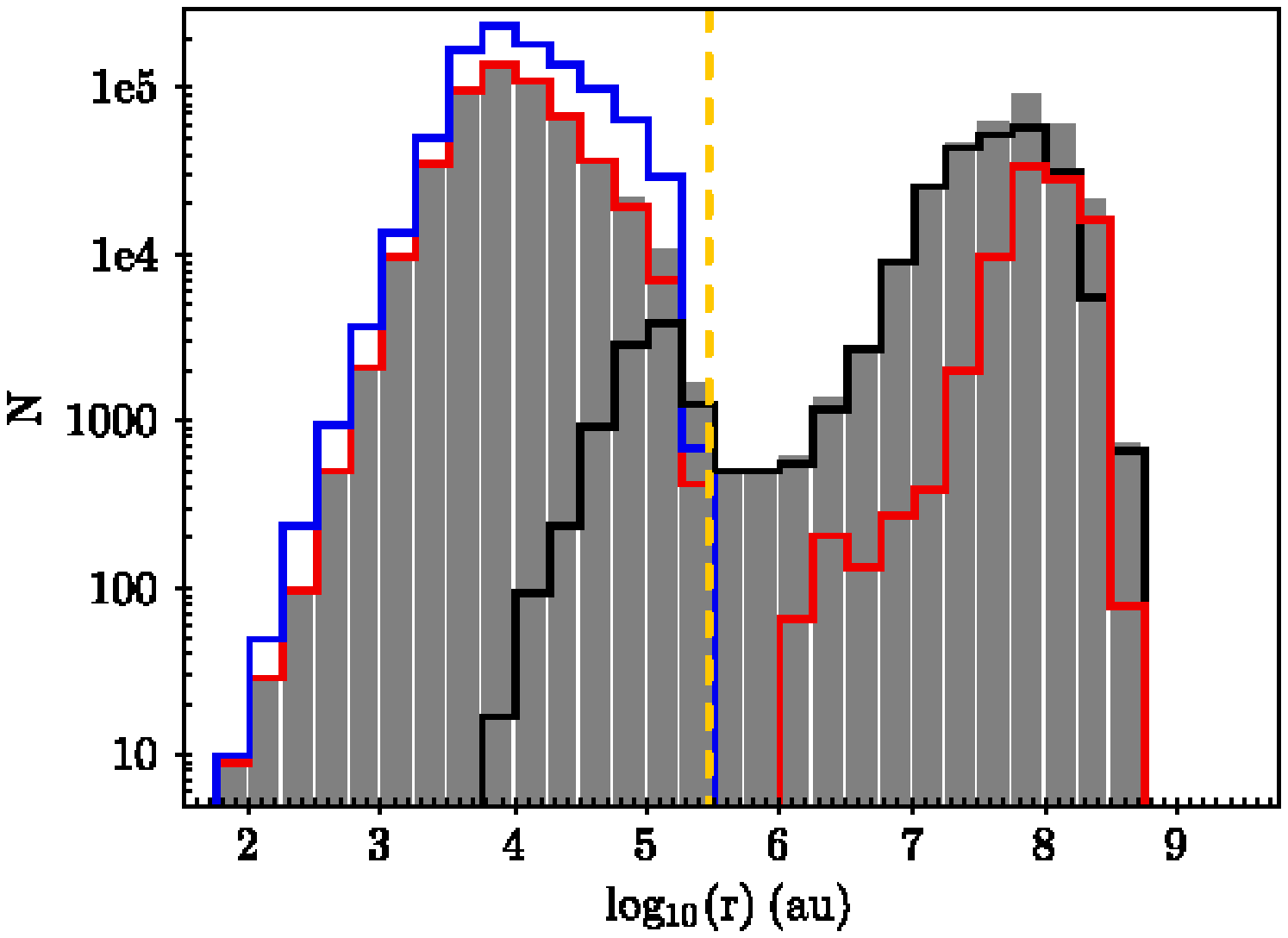}}
\subfigure{\includegraphics[width=0.99\columnwidth]{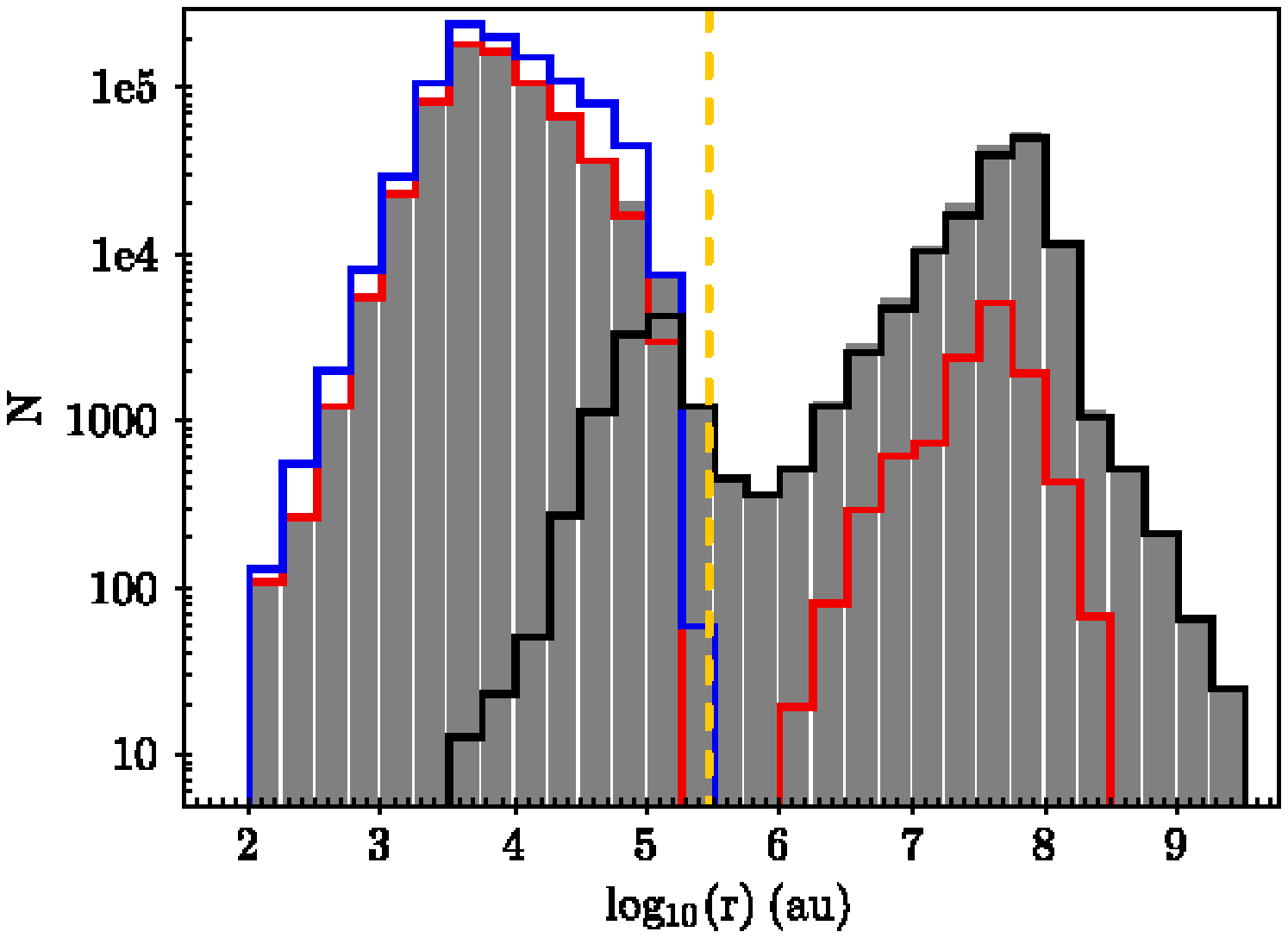}}
\subfigure{\includegraphics[width=0.99\columnwidth]{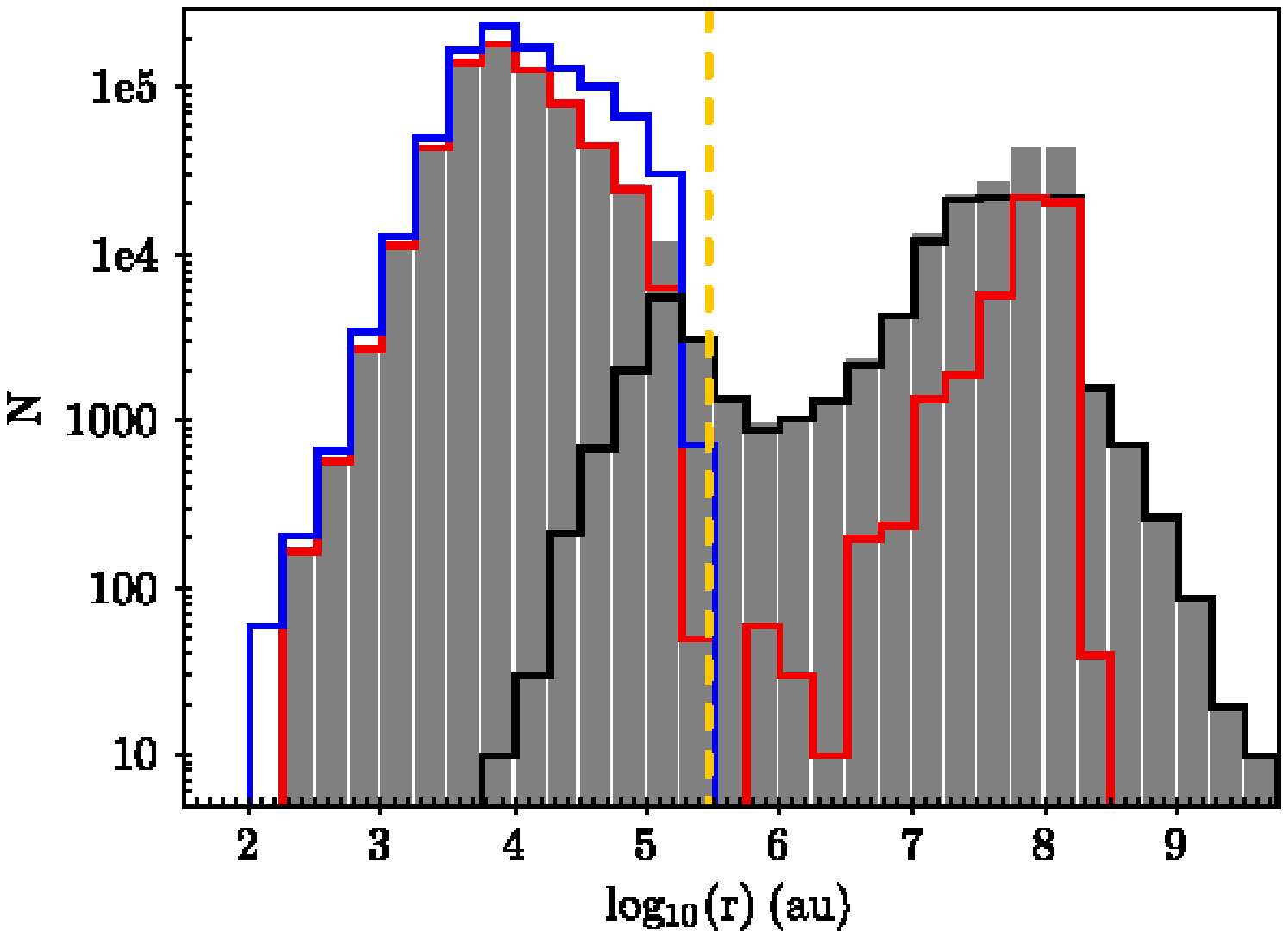}}
\caption{{\small Distribution of the final heliocentric distance of the particles of Samples 1 (top panel), 2 (middle panel) and 3 (bottom panel). The grey and blue histograms show the final and initial distributions. Moreover, we separate the final distribution in order to identify the 4 schematic regions of the Fig. \ref{fig1}: black and red histograms show particles with Jacobi constant lower and larger than the critical value $C_{crit}$, respectively, while dashed yellow line separate the particles with heliocentric distance lower and larger than $r_J$ (tidal radius), respectively. There is a minimum in the distribution at $\sim$ 10$^{6}$ au (3.6 $r_J$).}}
\label{fig3}
\end{figure}

In Fig. \ref{fig3} we show in grey a histogram for the final distribution of the heliocentric distances for particles that survive all the integration time. The Samples 1, 2 and 3 are shown in the top, middle and bottom panel, respectively, and we have included the initial distribution of each one in blue.  We can see the presence of two peaks, the first one for $r < r_J$ (or interior peak) which is due to the initial distribution. The exterior peak correspond to ejected particles ($r > r_J$) and is a consequence of the dynamic process beyond the tidal radius (see Sect. \ref{model}). These particles have an epicyclic motion in the $x$-$y$ plane with frequency $\kappa_G$, small oscillations in $z$ with frequency $\nu_G$, and a secular uniform motion in $y$, so this secular linear motion in the direction of rotation of the Galaxy plus the logarithmic representation of $r$ are responsible for the accumulation of particles at such large distance. Moreover, since the ejected particles have been moving away for less than 5 Gyr, they can not reach distances larger than a certain maximum determined by their velocity and the total integration time. The figure also shows a minimum in the number of objects at a heliocentric distance of $\sim$ 10$^{6}$ au ($\sim$ 3.6 $r_J$) which is also a consequence of the secular dynamic and the logarithmic representation.

To improve the dynamical analysis, it is possible to discriminate in the histogram shown in Fig. \ref{fig3} the particles belonging to the regions $B$, $U_1$, $U_2$ and $U_3$ indicated in Fig. \ref{fig1}, following the same color code.

We can see some similar characteristic in the three samples. There are more particles in the region $U_2$ than in region $U_3$, and the unbound particles of the region $U_1$ increase from $r \sim$ 10$^4$ au until some distance beyond 10$^5$ au  where they become dominant. Furthermore, we also find differences between the samples: for example, the Sample 1 has a greater percentage of particles in region $U_2$ (see, Table \ref{table1}), although the particles in this sample do not achieve distances larger than 10$^9$ au, while in Samples 2 and 3 this is possible for some objects. The Sample 2 has a small percentage of ejected objects in region $U_3$ in comparison with Samples 1 and 3, and there are a similar percentage of bound particles at the end of the simulation for Samples 2 and 3, while for Sample 1 this percentage is lower. In the next Section we try to explain why we observe such differences between the Samples.

\begin{figure}
\centering
\subfigure{\includegraphics[width=0.99\columnwidth]{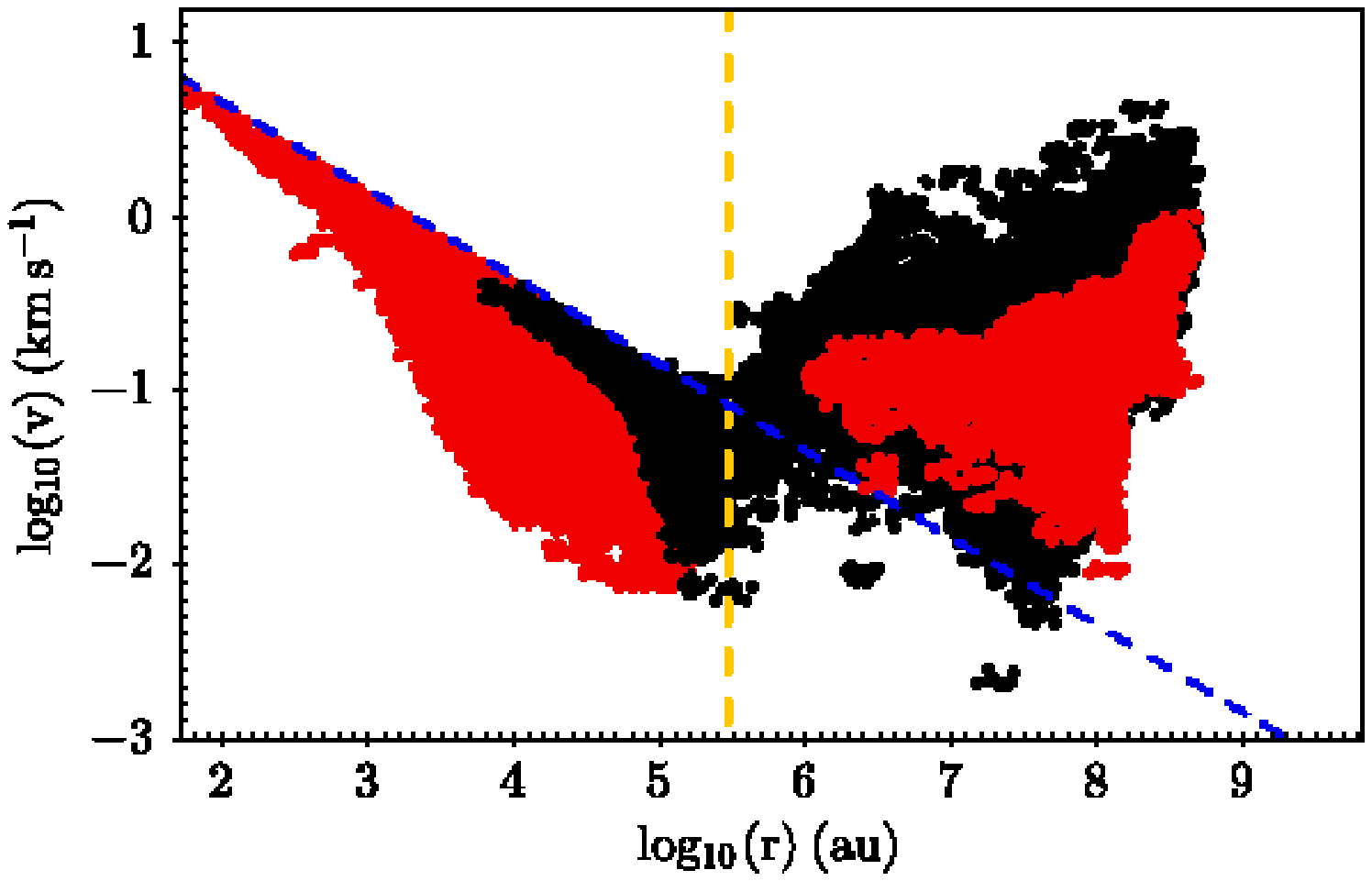}}
\subfigure{\includegraphics[width=0.99\columnwidth]{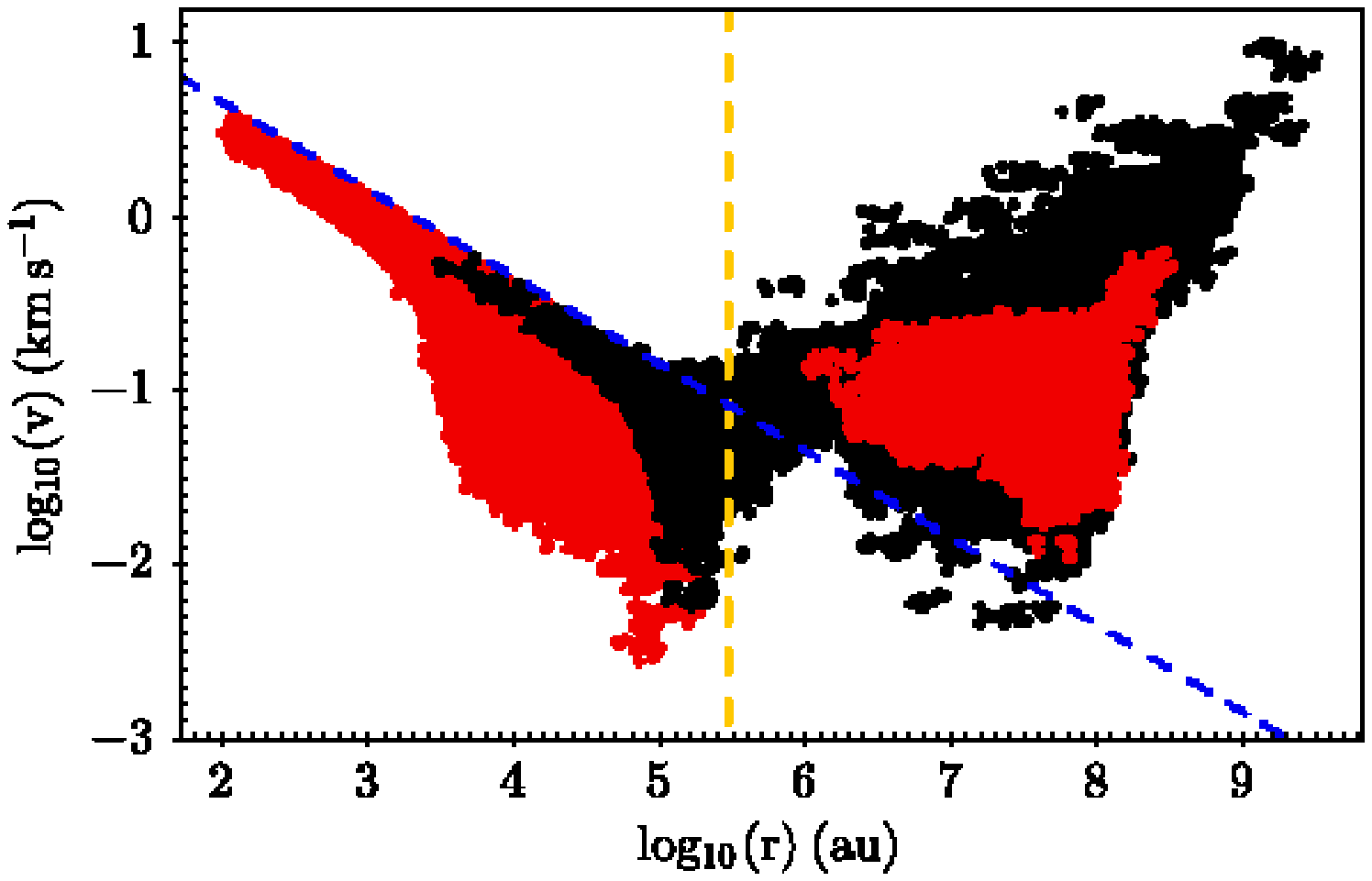}}
\subfigure{\includegraphics[width=0.99\columnwidth]{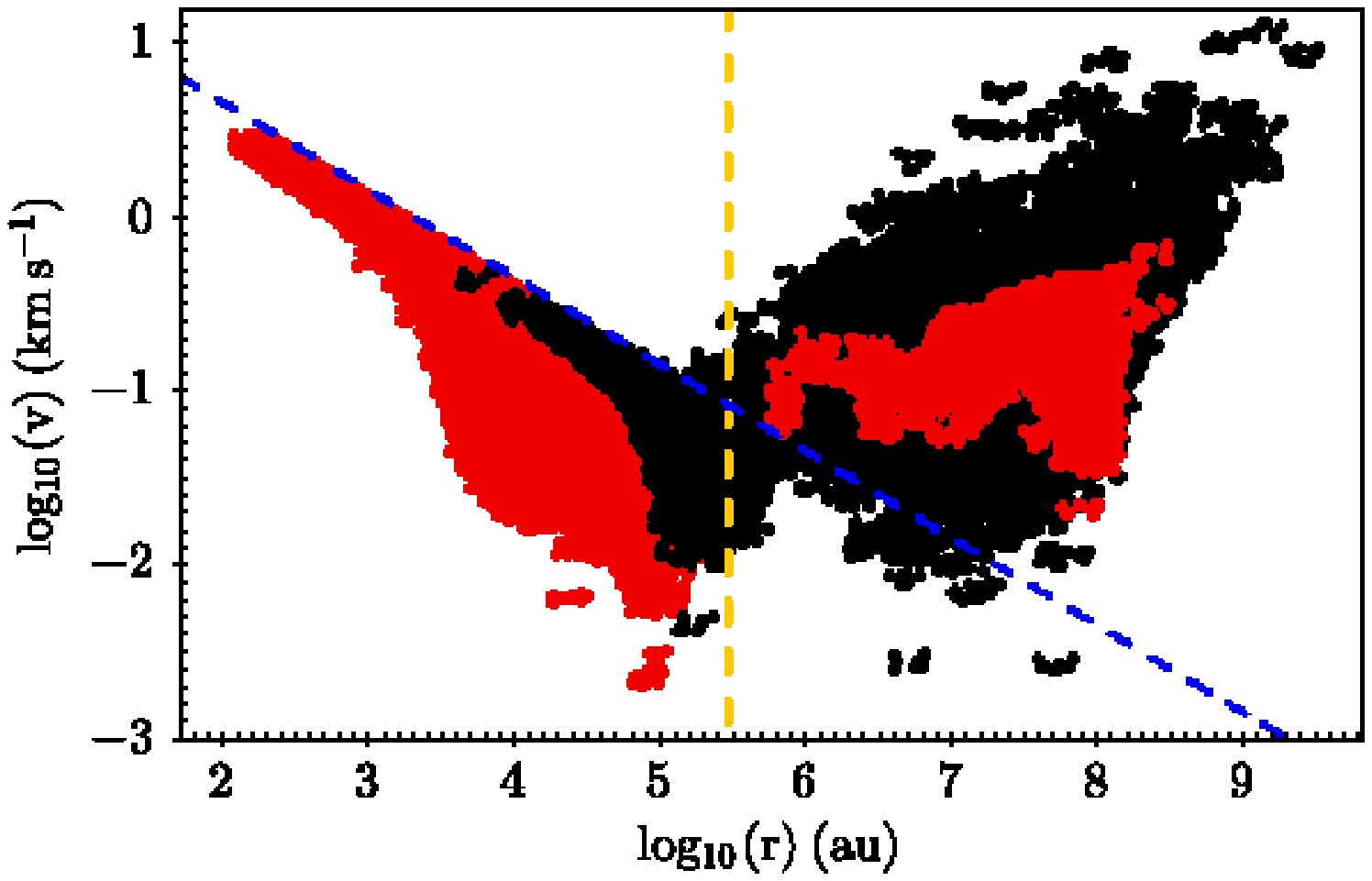}}
\caption{{\small Heliocentric distance vs. velocity at the end of simulation.  Particles with  value of $C_J$ higher than $C_{crit}$ are in black, while that particles with $C_J<C_{crit}$ are in red.  Blue and yellow dashed lines indicate the zero-energy limit for Keplerian orbits ($v_{E0}$) and the tidal radius ($r_J$), respectively.}}
\label{fig5}
\end{figure}

Fig. \ref{fig5} shows  the relation between the heliocentric distance and the velocity at the end of simulation for the Samples 1, 2 and 3 at the top, middle and bottom panel, respectively. The blue dashed line correspond to zero-energy for Keplerian orbits: $ v_{E0} = \sqrt{2 \mathcal{G} m_0 /r}$, while the tidal radius $r_J$ is indicated by a dashed yellow line. We have separated in each Sample the particles with $C_J < C_{crit}$, which are indicated in red, and particles with $C_J > C_{crit}$ in black, so that it is possible to appreciate the four schematic regions (see, Fig. \ref{fig1}).

For the three samples we found some similar characteristics in the final velocity distribution. The ejected particles ($r>r_J$) of region $U_2$ ($C_J > C_{crit}$) can reach values of maximum velocity larger than that of the particles of region $U_3$  ($C_J < C_{crit}$), which is because $v^2  \propto C_J$ (see, eq. \ref{eq4}). For the small percentage of  unbound particles close to the Sun (i.e., region $U_1$) we can see two groups, while particles with $r<$ 10$^5$ au live in a limited range of values of $v$, the objects with distances beyond 10$^5$ au have a greater dispersion. This is because the particles closer to the Sun are in a disturbed Keplerian regime of motion (Heisler \& Tremaine
1986; Correa-Otto et al. 2017), and due to the disturber effects of the Galactic potential they are unbound particles with high eccentric orbits which are close to the blue dashed line of Fig. \ref{fig5}. Instead, the unbound particles of region $U_1$ with $r>$ 10$^5$ au are strongly influenced by the Galactic tide, and therefore we can find them even at low values of $v$.

Besides, it is possible to find some differences in the final distribution of velocities of the three samples, specially between Sample 1 and the other two. The main difference is the maximum value of $v$ achieved by the particles of region $U_2$ in Samples 2 and 3 that can reach 10 km s$^{-1}$, a value which is not observed in Sample 1. This difference explain why for the Samples 2 and 3 we can observe particles of the region $U_2$ at distances larger than 10$^9$ au, while the corresponding particles of Sample 1 do not achieve such distances.

On the other hand, in Fig. \ref{fig5} at $r \sim r_J$ it is possible to see that although there are some unbound particles with $v > v_{E0}$ (i.e., hyperbolic orbits with $C_J > C_{crit}$), most of the unbound objects in such region live below the blue line in elliptic orbits with $C_J > C_{crit}$. In the case of ejected particles ($r > r_J$), it is worth to note that the unbound objects of region $U_3$ are over the theoretical zero-energy line for Keplerian orbits, which is in agreement with a hyperbolic orbit. Instead, there are some particles of region $U_2$ that live under the dashed blue line of zero velocity. These are interesting results, because such group of particles with $v < v_{E0}$ have been ejected from the Solar System, but according to the keplerian two-body problem they are still bound to the Sun. So, these results confirm the need to consider the Galactic potential for studies about the dynamic evolution of the outer region of the Oort Cloud.

Finally, it is worth to note that the final distributions of $r$ and $v$ are similar to that of Jiang \& Tremaine (2010) for wide binary stars. Such result indicates that there is not a dynamic difference between a disturbed two-body problem (i.e., star-star) and a disturbed restricted two-body problem (i.e., Sun-comet), when the Galaxy is the disturber, which is in agreement with the result obtained by Correa-Otto et al. (2017) and confirm that the dynamical evolution of a pair of objects in the Galaxy is slightly dependent on the mass of the binary system.

\subsection{Evolution of the ejected particles}\label{evol}

\begin{figure}
\centering
\subfigure{\includegraphics[width=0.99\columnwidth]{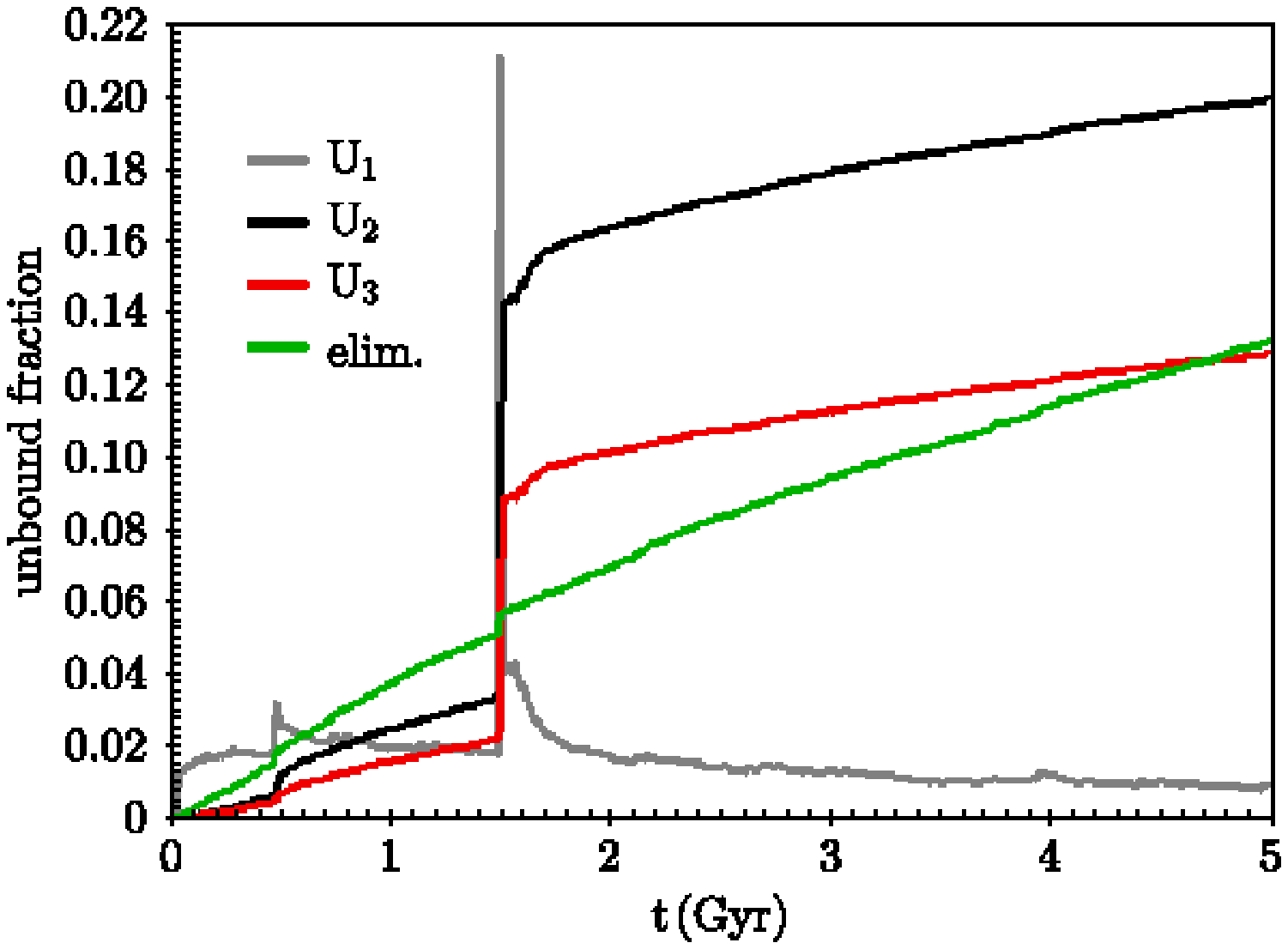}}
\subfigure{\includegraphics[width=0.99\columnwidth]{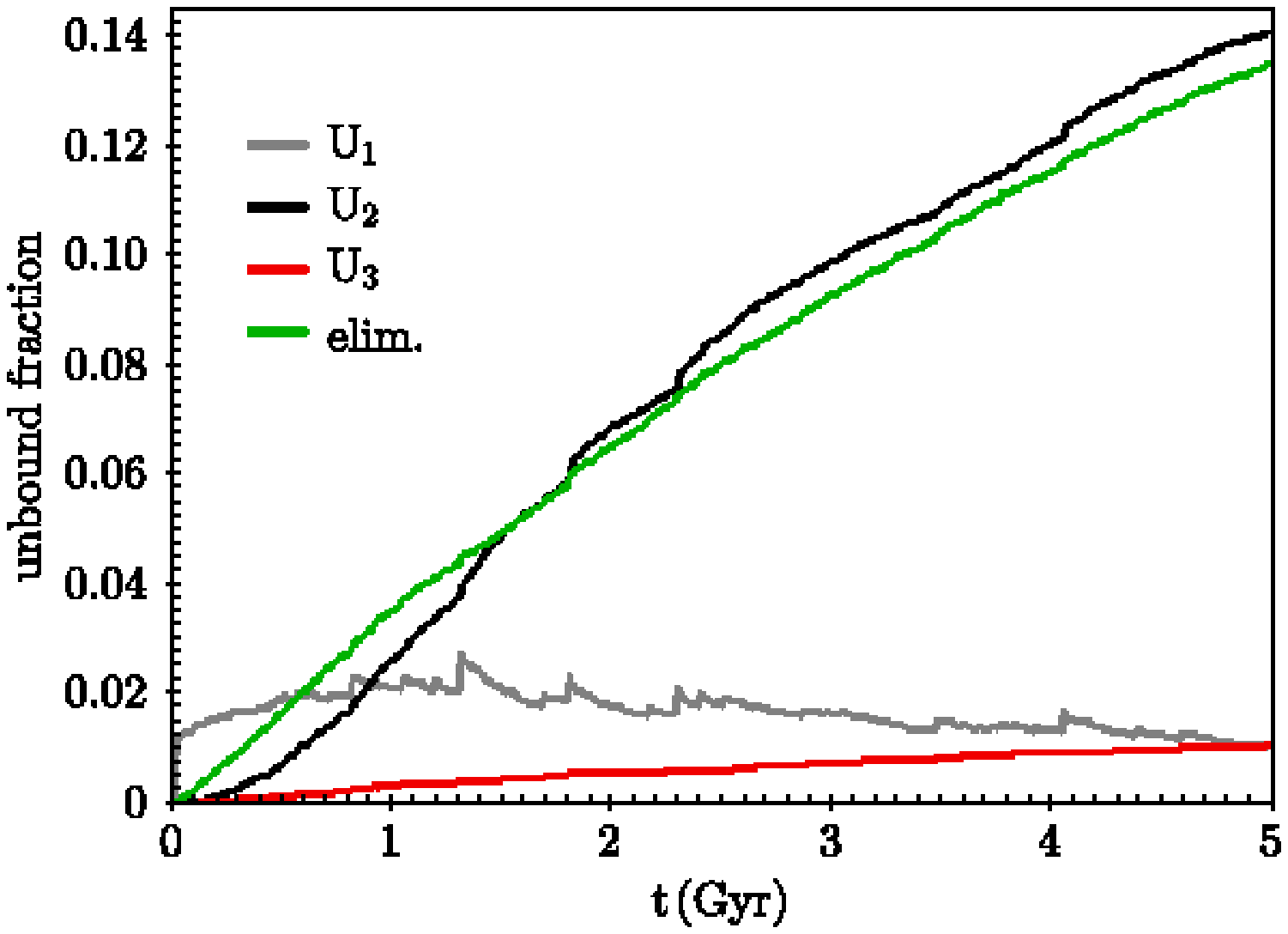}}
\subfigure{\includegraphics[width=0.99\columnwidth]{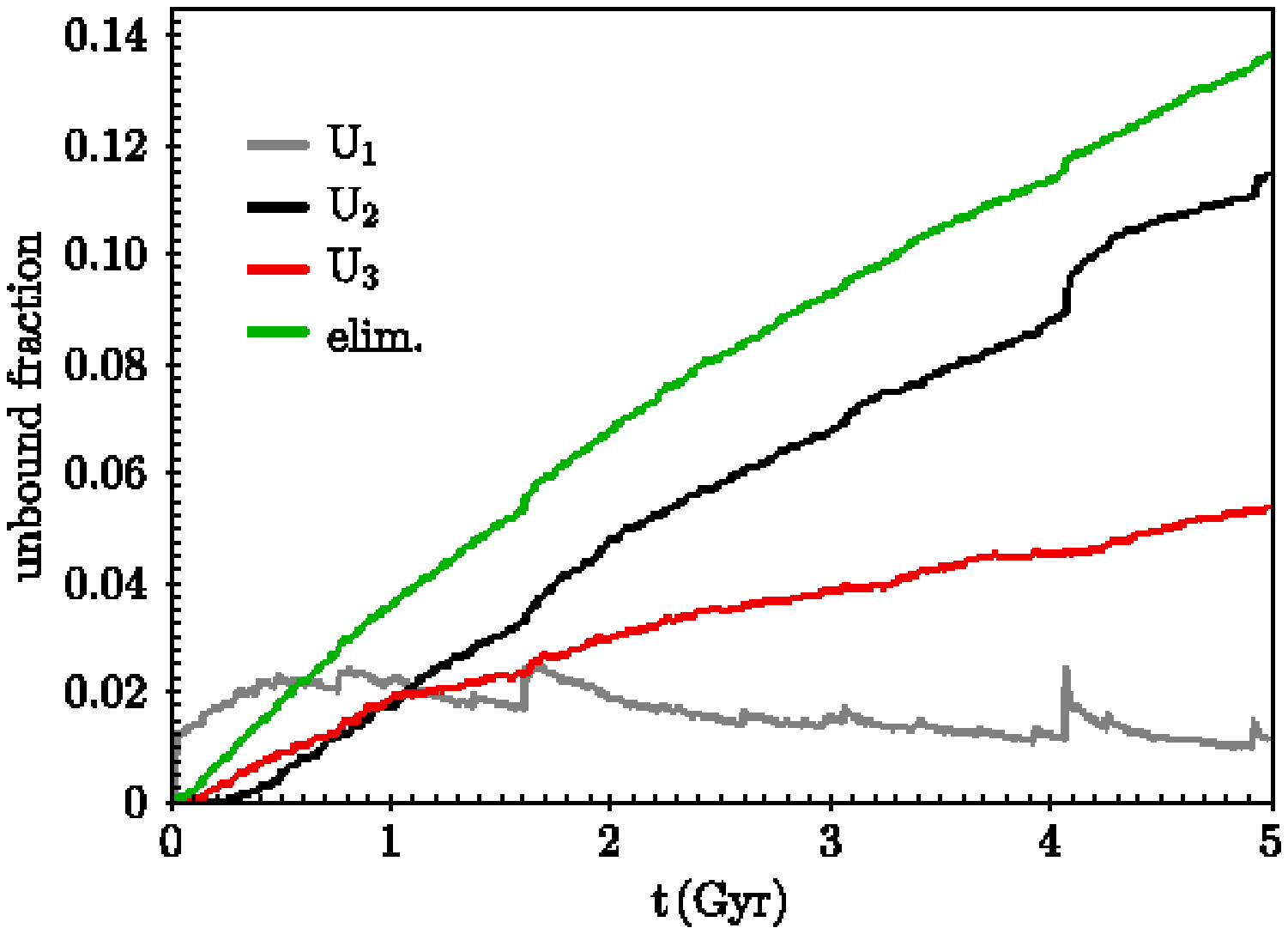}}
\caption{{\small Evolution of the fraction of unbound particles separated according to the schematic regions $U_1$ (grey line), $U_2$ (black line)  and $U_3$ (red line), for Samples 1 , 2  and 3 in top, middle and bottom panel, respectively. Moreover, we have also include the evolution of the fraction of eliminated particles in green line. }}
\label{fig7}
\end{figure}

To improve our comprehension about the dynamical process that eject particles from the Solar System, we analyse the temporal evolution of the particles that ends the integration in an unbound state. Fig. \ref{fig7} shows the evolution of these particles, separated according to the defined regions $U_1$, $U_2$ and $U_3$, for all the samples. The particles in region $U_1$ shown the same behaviour for the three samples, with an increment until 0.02 at the beginning of the simulation and a stabilization at the end close to 0.01. The Sample 1 however shows an important instantaneous increase to 0.21 at $\sim$ 1.5 Gyr, which quickly decreases 0.2 Gyr later. For the particles of regions $U_2$ and $U_3$, in all the samples we can see a continuous increase of the fraction of unbound ejected particles, and there is an important increase of ejected particles of Sample 1 at 1.5 Gyr.

\begin{figure}
\centering
\subfigure{\includegraphics[width=0.99\columnwidth]{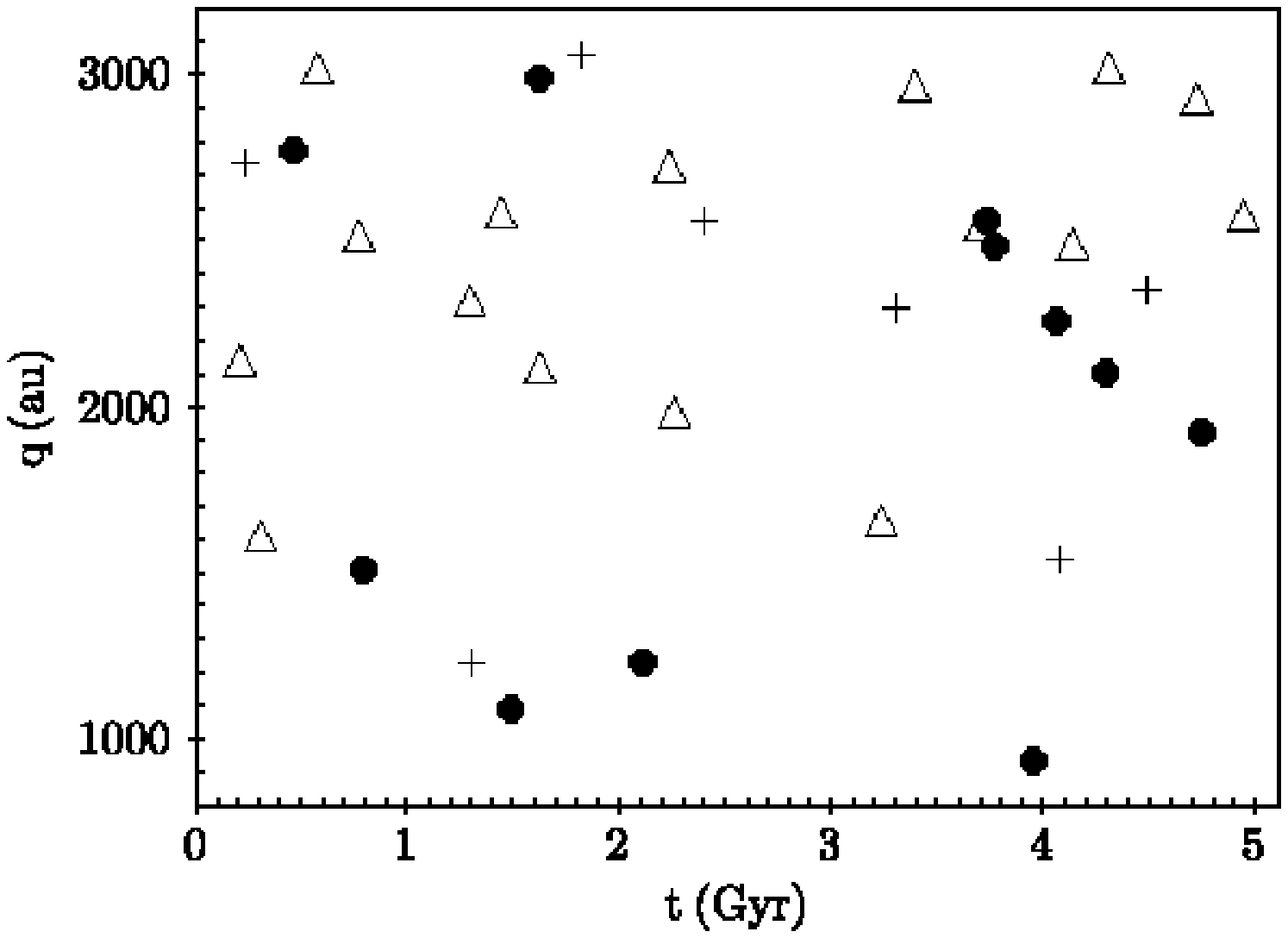}}
\subfigure{\includegraphics[width=0.99\columnwidth]{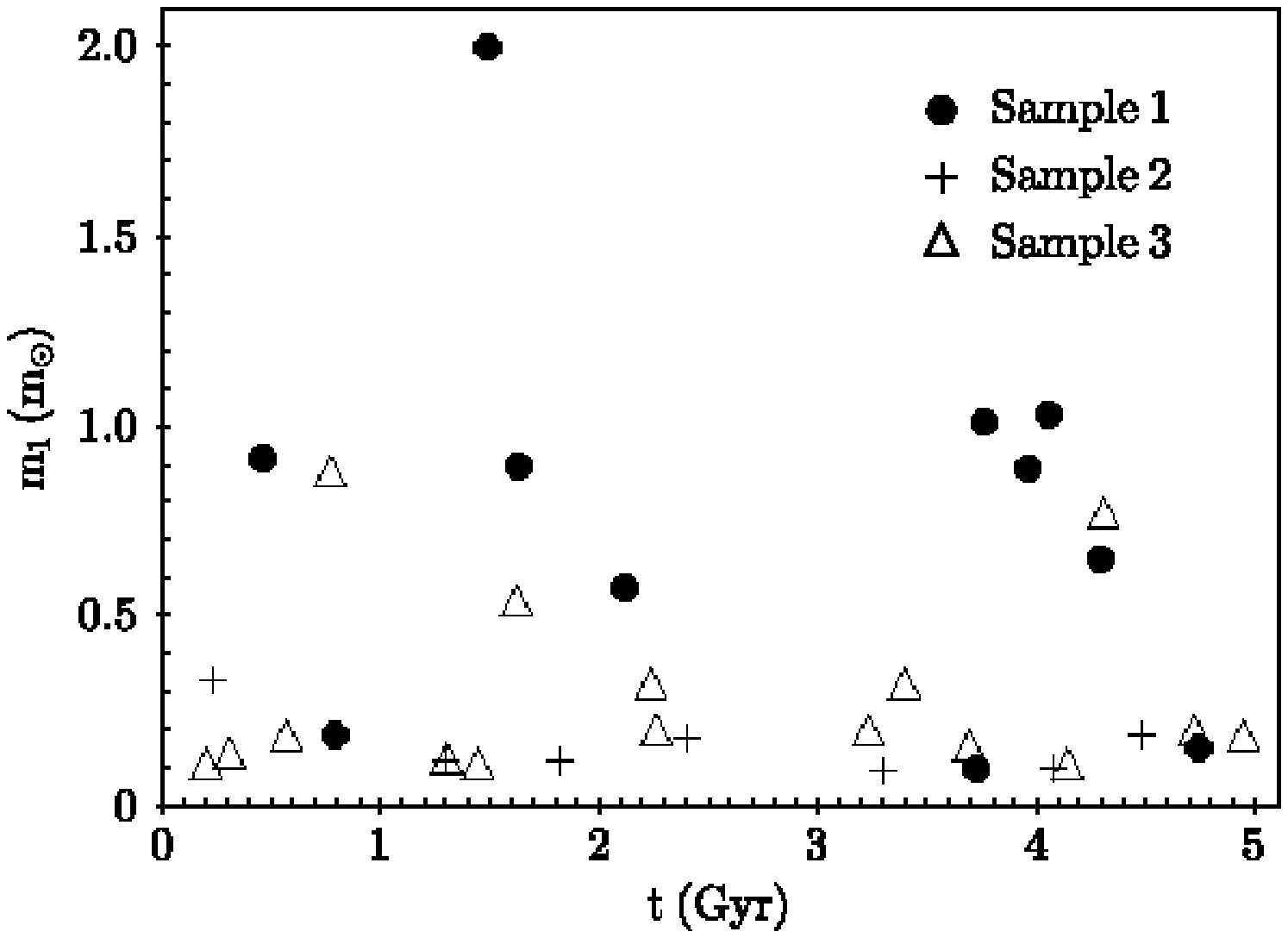}}
\subfigure{\includegraphics[width=0.99\columnwidth]{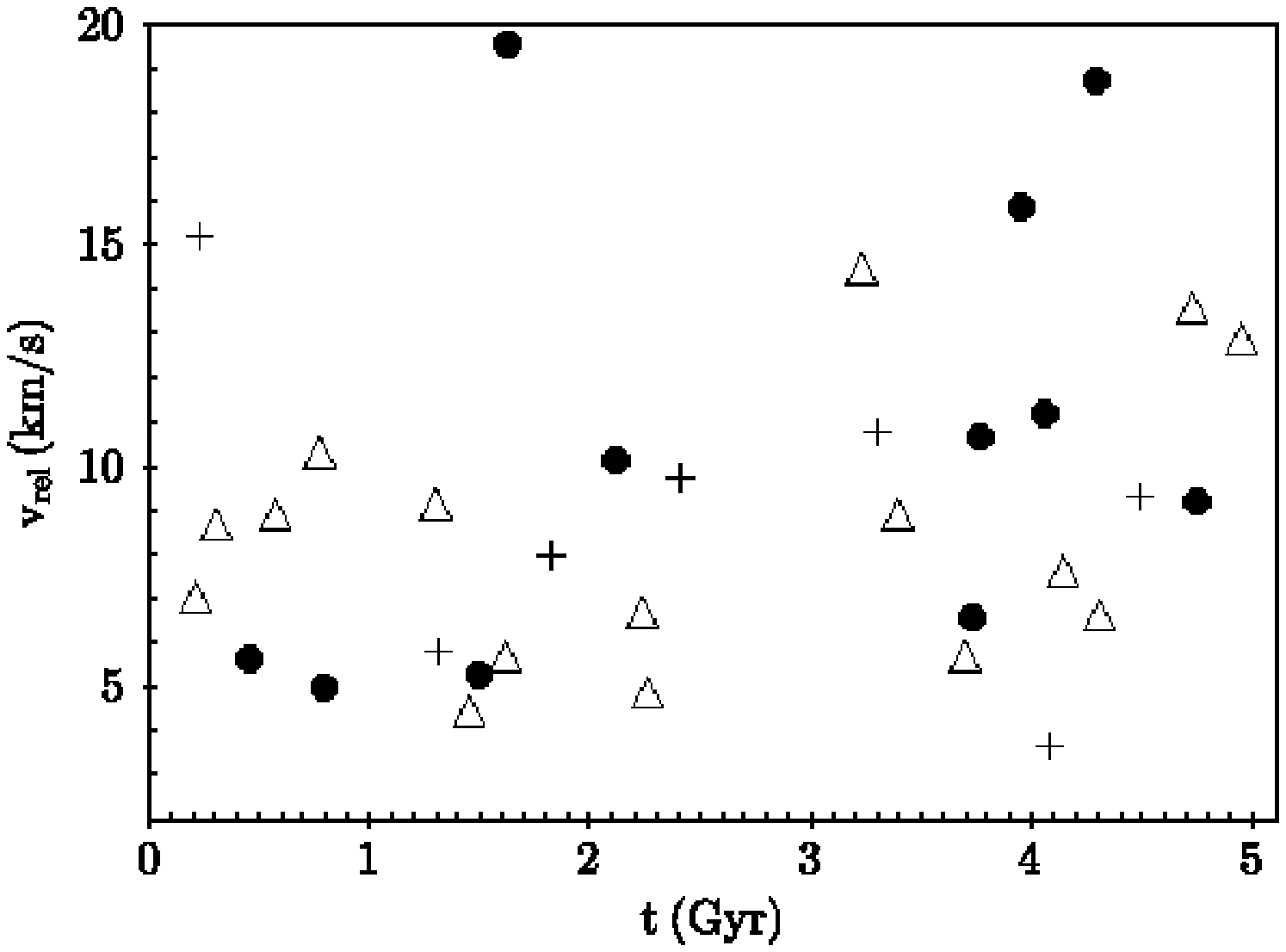}}
\caption{{\small Dynamic characteristic of the closest stellar passages ($q<$ 3000 au). Top panel: impact parameter. Middle panel: stellar mass. Bottom panel: initial velocity. Full circle, cross and unfill triangles  correspond to the samples 1, 2 and 3, respectively. }}
\label{fig8}
\end{figure}

The ejection of particles is a consequence of the stellar passages whose influence can be appreciated in the peaks of unbound particles in region $U_1$ and in stepped increases of ejected particles of regions $U_2$ and $U_3$. Some stellar passages can produce important consequence in the process of ejection, for example the important increase at $\sim$ 1.5 Gyr of Sample 1. The influence in the Cloud of a stellar passage depends on its dynamical characteristic. In order to identify the most important stellar passages of each sample, in the three panels of Figure \ref{fig8} are shown the impact parameter ($q$),  the stellar mass of the disturbing star ($m_1$) and the relative velocity ($v_{rel}$) for all the stellar passages with  $q<$ 3 $\times$ 10$^3$ au and for the three samples. For a maximum impact parameter $q_M \sim$ 3 $\times$ 10$^3$ au, the predicted number of passages is $\sim$ 9 (see Sect. \ref{model}), and the results for the three samples are in agreement with the theoretical prediction because we find 11, 7 and 16 stellar passages for the Samples 1, 2 and 3, respectively.

Then, for all the samples we find matches between some of the stellar passages in Fig. \ref{fig8} and the instant when the number of unbound particles increase for region $U_1$ in Fig. \ref{fig7}. The two most important events for each Sample take place at $\sim$ 0.5 and 1.5 Gyr for the Sample 1, at $\sim$ 1.3 and 1.8 Gyr for Sample 2, and  at $\sim$ 1.6 and 4.1 Gyr for Samples 3. Of these six stellar passages, five of them produce an increase lower than 0.01 in the fraction of unbound particles of region $U_1$, while the relative velocity seems to be the parameter that defines the impact in the dynamical evolution of the cloud of particles, since all them have $v_{rel}<$ 8 km/s. Instead, the stellar passage of Sample 1 which happens at $\sim$ 1.5 Gyr has important consequences in the evolution of the synthetic sample of particles. The corresponding star has a mass of $\sim$ 2 m$_\odot$ and a small relative velocity ($\sim$ 5 km/s), and it match with the instants of most important stepped growth in the fraction of unbound particles (see Fig. \ref{fig7}). It is worth to note that the high increase of particles in region $U_1$ ($\sim$ 0.18) at 1.5 Gyr is later distributed between the two regions of ejected particles, $\sim$ 0.11 in region $U_2$ and $\sim$ 0.07 in region $U_3$, in a way such that at $\sim$ 1.7 Gyr the fraction of particles in the region $U_1$ returns to the normal value (0.01). Then, while in the other samples the most important stellar passages eject a fraction of particles that does not exceed $\sim$ 0.01, the exceptional event of Sample 1 can duplicate the total fraction of ejected particles in a relative short time ($\sim$ 0.2 Gyr).  Here after, we will call this stellar passage so efficient to eject particles from the Solar System an "special event".

In the six selected cases of the previous paragraph we can see that the increase in the unbound fraction is always appreciated before in region $U_1$, and an instant later in regions $U_2$ and $U_3$. This behaviour seems to indicate a two-step mechanism to eject particles from the system. First, the particles are excited by an increment of the Jacobi constant due to a stellar passage, which moves the particles from region $B$ to region $U_1$. The second step is an increase of the heliocentric distance produced by the following passages plus the Galactic tide, but it could happen in two different ways: with a small change of $C_J$ where the particle ends in the region $U_2$, or with an important decrease of $C_J$ where the particle ends in the region $U_3$. This mechanism is shown schematically in Fig. \ref{fig1}, where the green star represent the initial position of a fictitious particle, which with a first stellar passage jump to the region $U_1$ (unfilled green star 1) by an increase of $C_J$. Then, the particle can be moved to the region $U_2$ (unfilled green star 2) by the Galactic potential or by other stellar passage, or to the region $U_3$ (unfilled green star 3), but in this last case only a stellar passage can move the particle with a decrease of $C_J$. We call to this mechanism "continuous process" for the continuous ejection of particles out of the Solar System.

Moreover, our final results (see, Table \ref{table1}) show that the ejection of particles to the region $U_2$ is the most probable, which can be explained easily through the method described in the previous paragraph. Then, to move particles from region $U_1$ to region $U_3$ we need a change in $C_J$ for which we need a stellar passage. On the other hand, the region $U_2$ is the natural destination for a unbound particle because the Galactic potential can eject it from the region $U_1$ even without a stellar passage.

Another result is obtained when we compare Samples 2 and 3 where the percentage of unbound particles is similar, but we can see that the number of ejected particles into regions $U_2$ and $U_3$ is different for both samples. Such result indicates that there is a stochastic process filling the regions of ejection which is associated to  the effect produced by the stellar passages. Then, we have find that the sequence of stellar passages modulate the dynamic mechanism of escape from the Solar System.

Heisler \& Tremaine (1986)shown that the stellar passages are responsible of the comet showers, but our results show that they are also responsible for the ejection of particle from the Solar System. In Fig. \ref{fig7} we also show the dynamical evolution of the eliminated particles which also has a linear increase like that of the ejected particles.

The mechanism of "continuous process" ejects and injects material from the Cloud, which is particularly confirmed in Sample 2 where there are a small fraction of particles in region $U_3$ and we can see a similar increase rate of the injected and ejected particles (green and black lines in Fig. \ref{fig7}). Also, for Samples 2 and 3 the final fraction of unbound particles ($\sim$ 0.17) is similar to that of eliminated particles ($\sim$ 14 \%), which indicates that the "continuous process" has the same efficiency to eject or inject particles from the Oort Cloud. Moreover, while the "special event" of Sample 1 can duplicate the number of ejected particles, it does not have important consequences for the total number of eliminated particles. This difference indicate that the "special event" is very efficient to unbound particles, but it is not so efficient to  make particles cross the threshold of 35 au.

 \begin{figure}
% \begin{center}
%\epsfig{figure=maximo_evolucion_tot.eps,width=0.99\columnwidth}
\centering
\subfigure{\includegraphics[width=0.99\columnwidth]{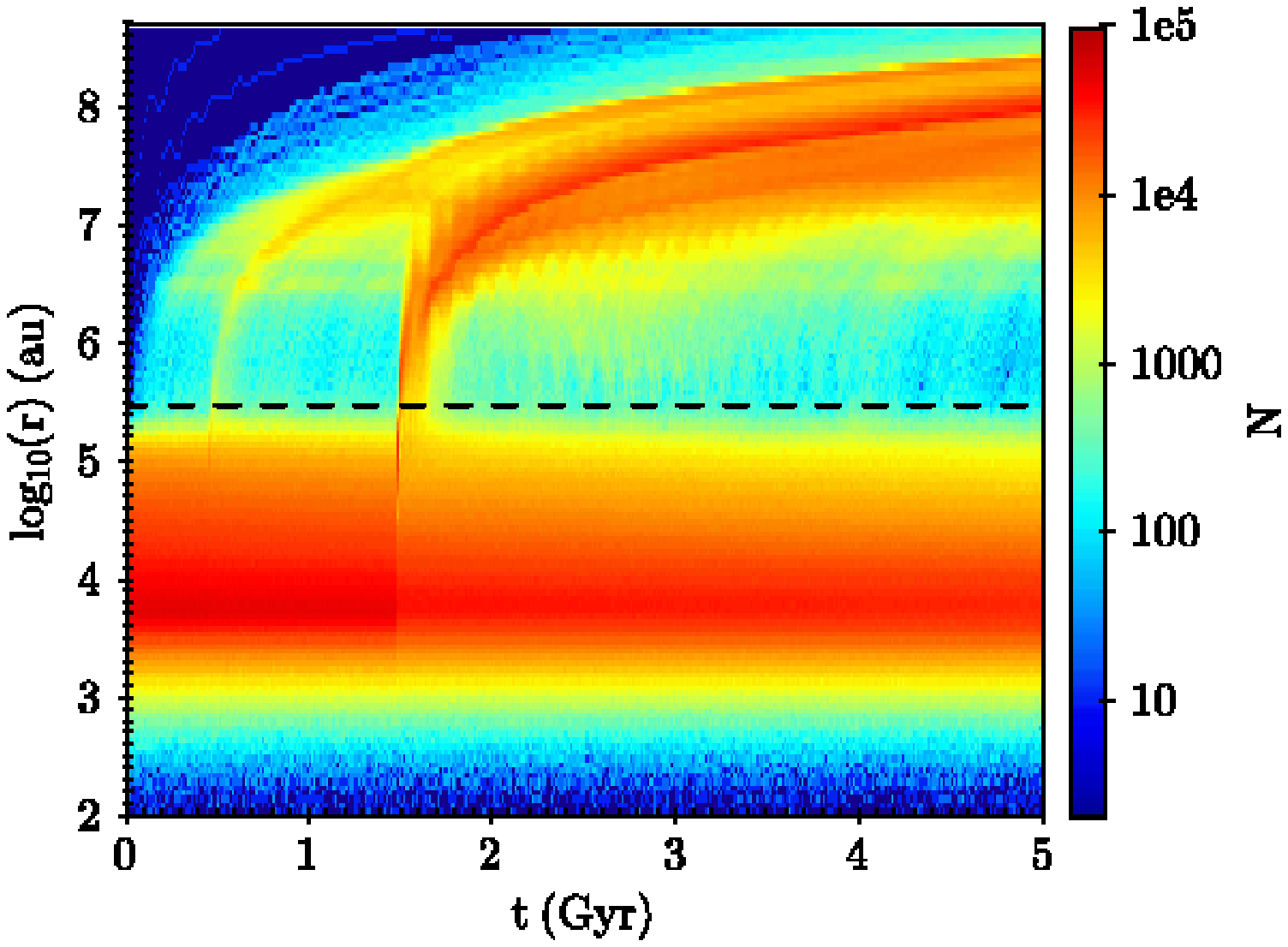}}
\subfigure{\includegraphics[width=0.99\columnwidth]{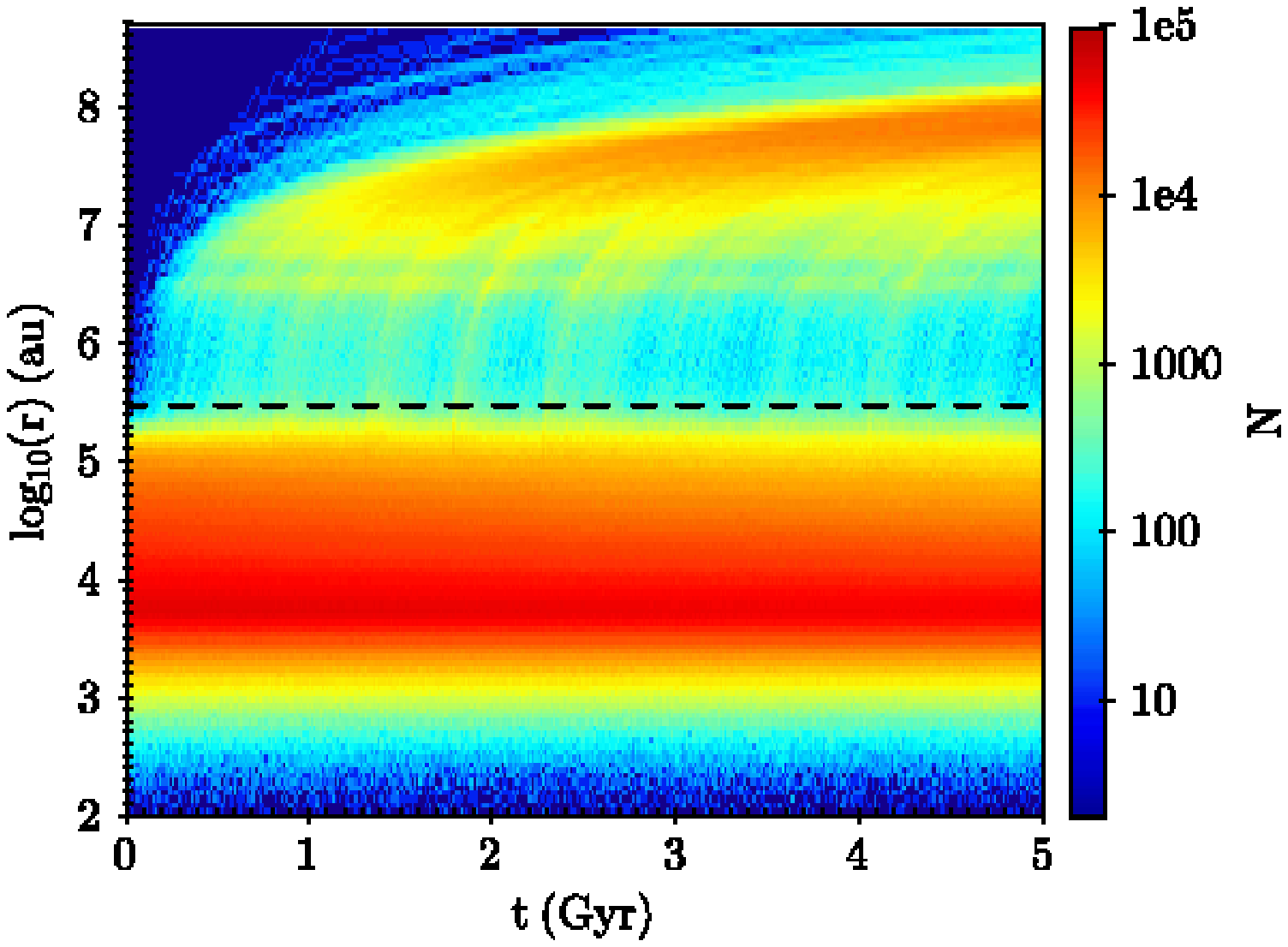}}
\subfigure{\includegraphics[width=0.99\columnwidth]{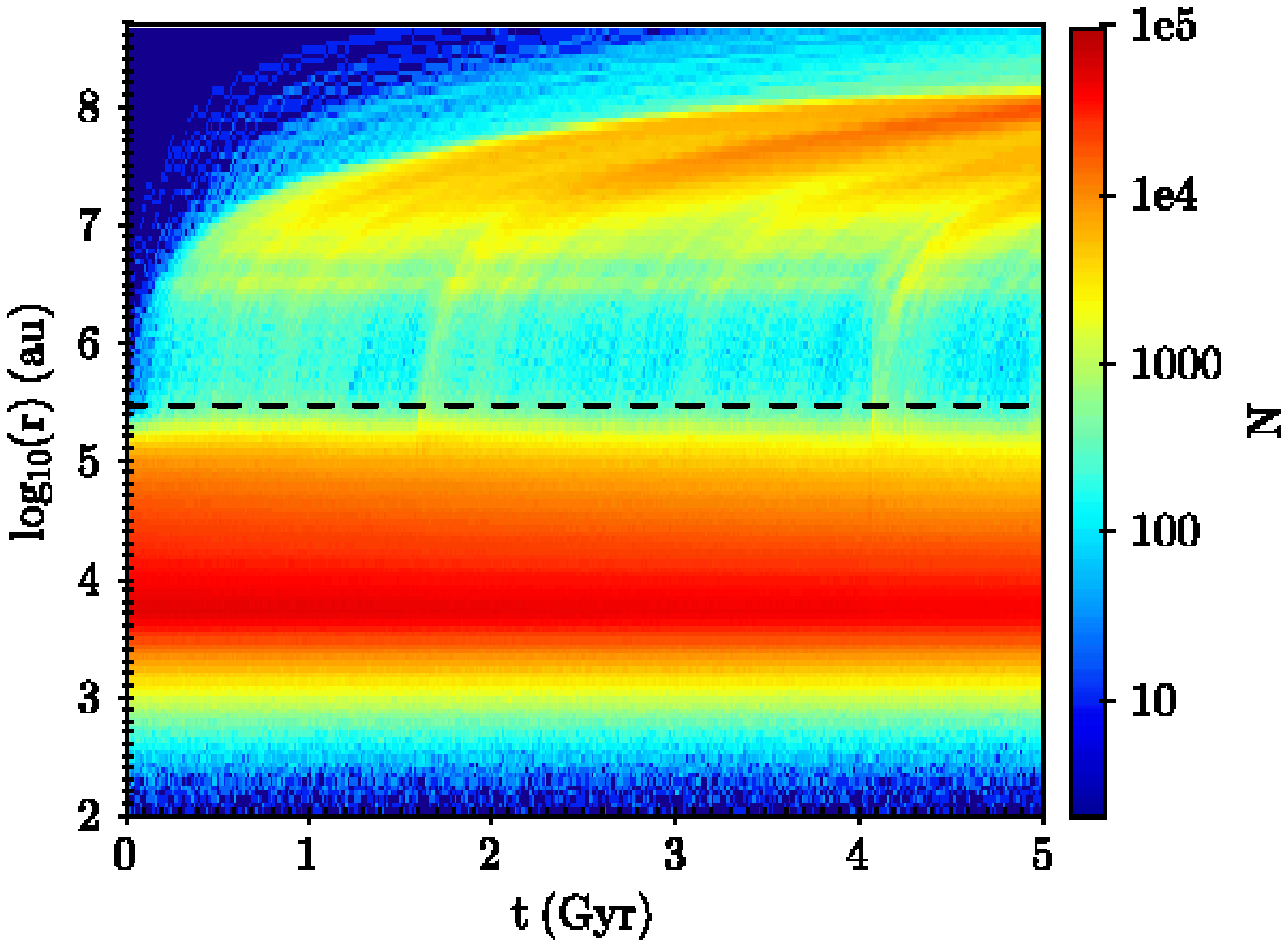}}
\caption{{\small Evolution of the distribution of the heliocentric distance $r$. The tidal radius is shown with a dotted black line. From top to bottom, Samples 1, 2 and 3.}}
% \end{center}
\label{fig9}
\end{figure}

Moreover, if in Sample 1 we remove the fraction of ejected particles due to the "special event" ($\sim$ 0.18), we would have only a fraction of $\sim$ 0.16 of particles ejected, which is similar to the other two samples. Then, this allows us to arrive at an important result: in absence of a "special event" the final percentage of ejected particles is independent of the sequence of stellar passages.

Finally, to complete our analysis we show the evolution of the density distribution for the heliocentric distance of Sample 1, 2 and 3 in the top, middle and bottom panel of Figure \ref{fig9}, respectively. The tidal radius $r_J$ is indicated with a dotted black line. The effects produced by the stellar passages to eject particles from the Oort Cloud are evident in the plots. It is possible to see that the objects are ejected as a shower of particles in a similar way to that observed for the cometary showers. As we have described in Sect. \ref{model}, every ejected particle has a linear increase of the heliocentric distance (see, eq. \ref{eq8}), so a set of particles ejected by the same stellar passage will evolve together towards high values of $r$ with small differences in the fast epicyclic motion on the $x$-$y$ plane. For example, with the great number of particles ejected by the "special event" at 1.5 Gyr in Sample 1, we can confirm the secular evolution due to the Galactic potential described in Sect. \ref{model}. Moreover, this bulk of high density seems to increase with time, which is because our logarithm representation of $r$, as we have explained in Sect. \ref{result}.  Therefore, the temporal evolution of the distribution of $log_{10}(r)$ presented in Fig. \ref{fig9} allows us to see the formation of the apparent exterior peak at $\sim$ 10$^8$ au for the three samples, which confirms our arguments of Sect. \ref{result}. However, our results about the evolution of the ejected particles are limited by our approximations, which will be detailed in the next section. On the other hand, we can see in Fig. \ref{fig9} for the three samples, that with the temporal evolution the successive stellar passages decrease the initial density of the Oort Cloud. Hence, the quantity of material to be ejected decrease too, which explain why the blue region around the tidal radius increase with the time. This result completes our explanation of Sect. \ref{result} about the presence of a minimum in density at $\sim$ 3.6 $r_J$.

\subsection{The unbound Oort Cloud}

\begin{figure}
\centering
\subfigure{\includegraphics[width=0.48\columnwidth]{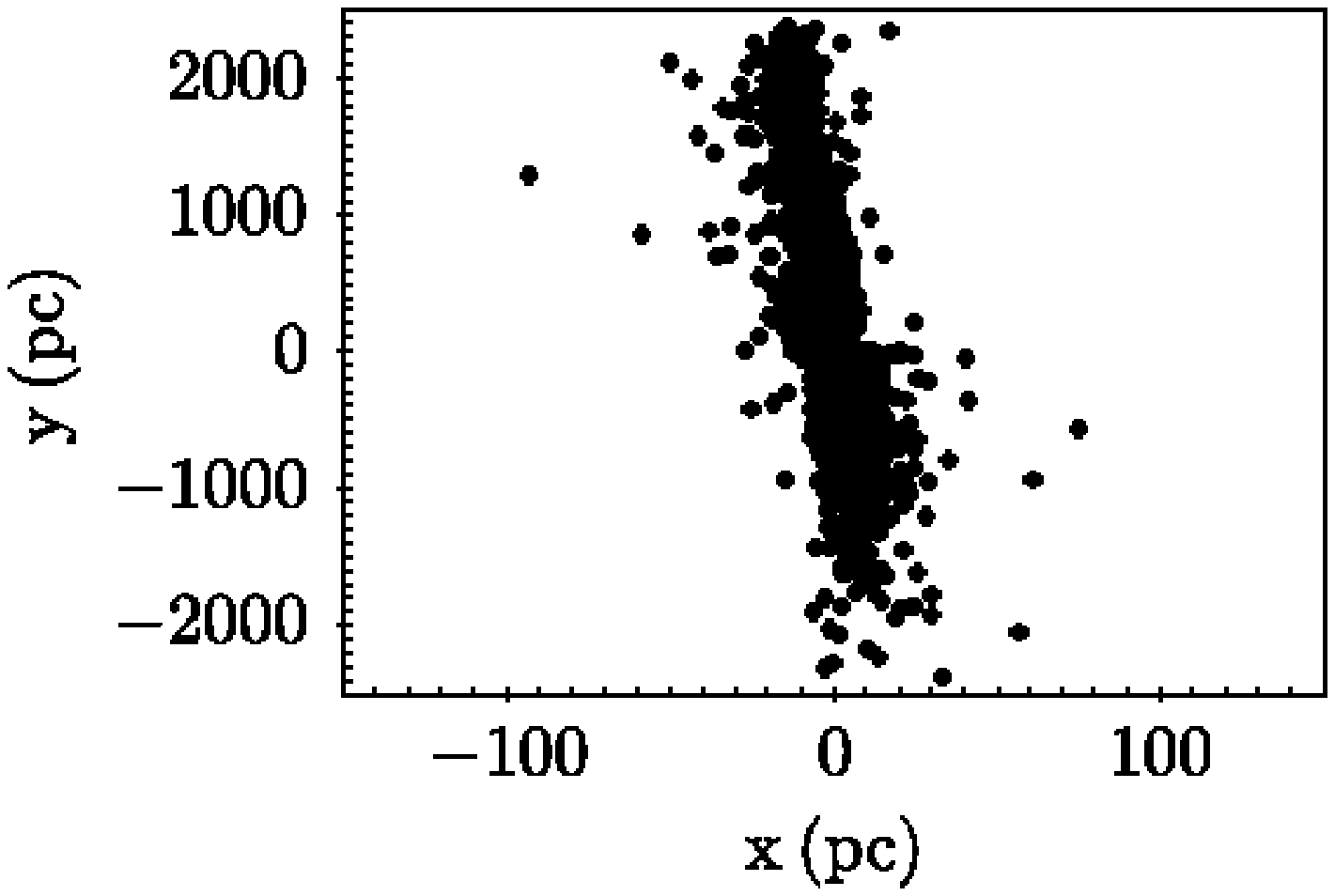}}
\subfigure{\includegraphics[width=0.48\columnwidth]{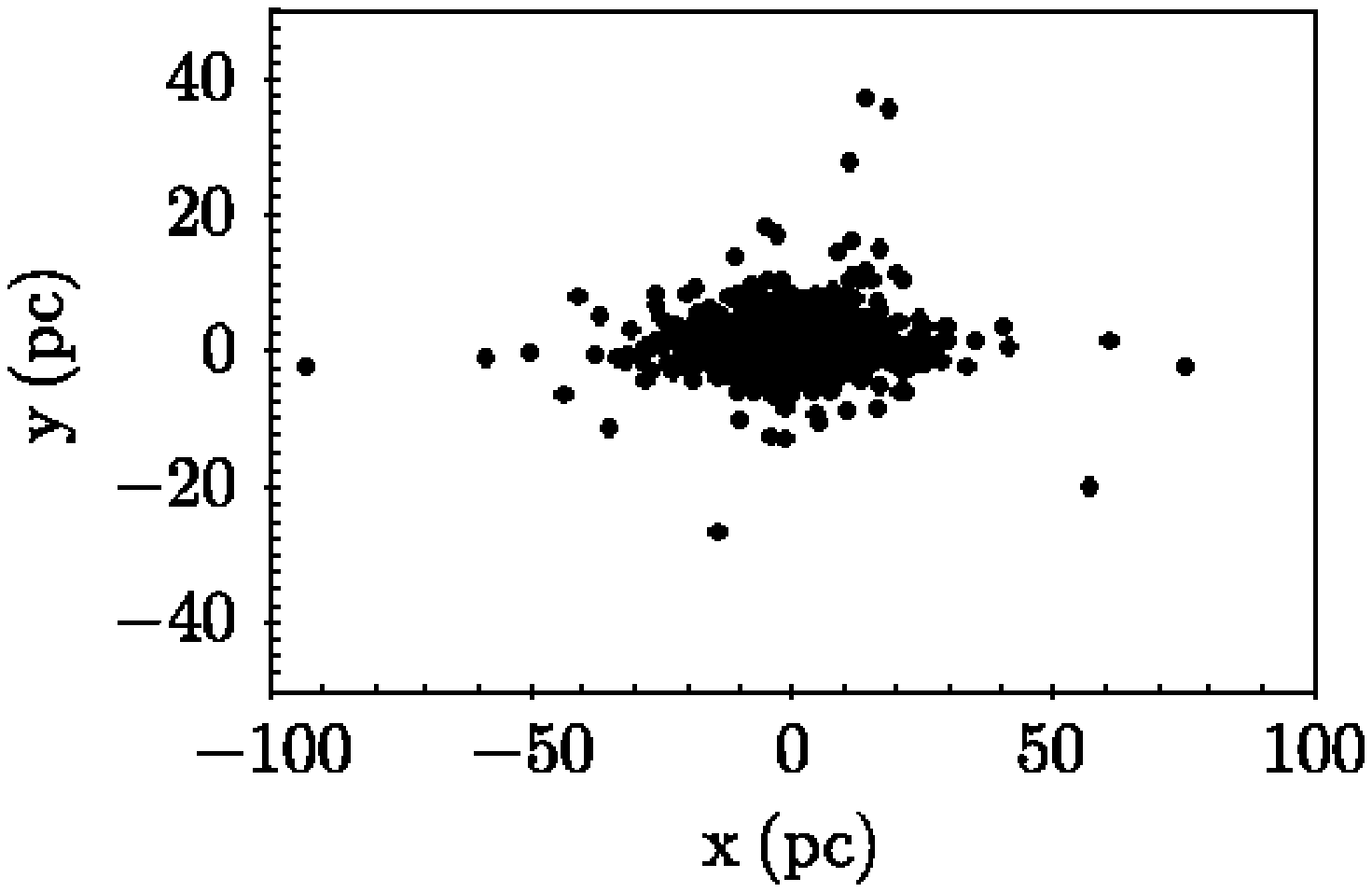}} \\
\subfigure{\includegraphics[width=0.48\columnwidth]{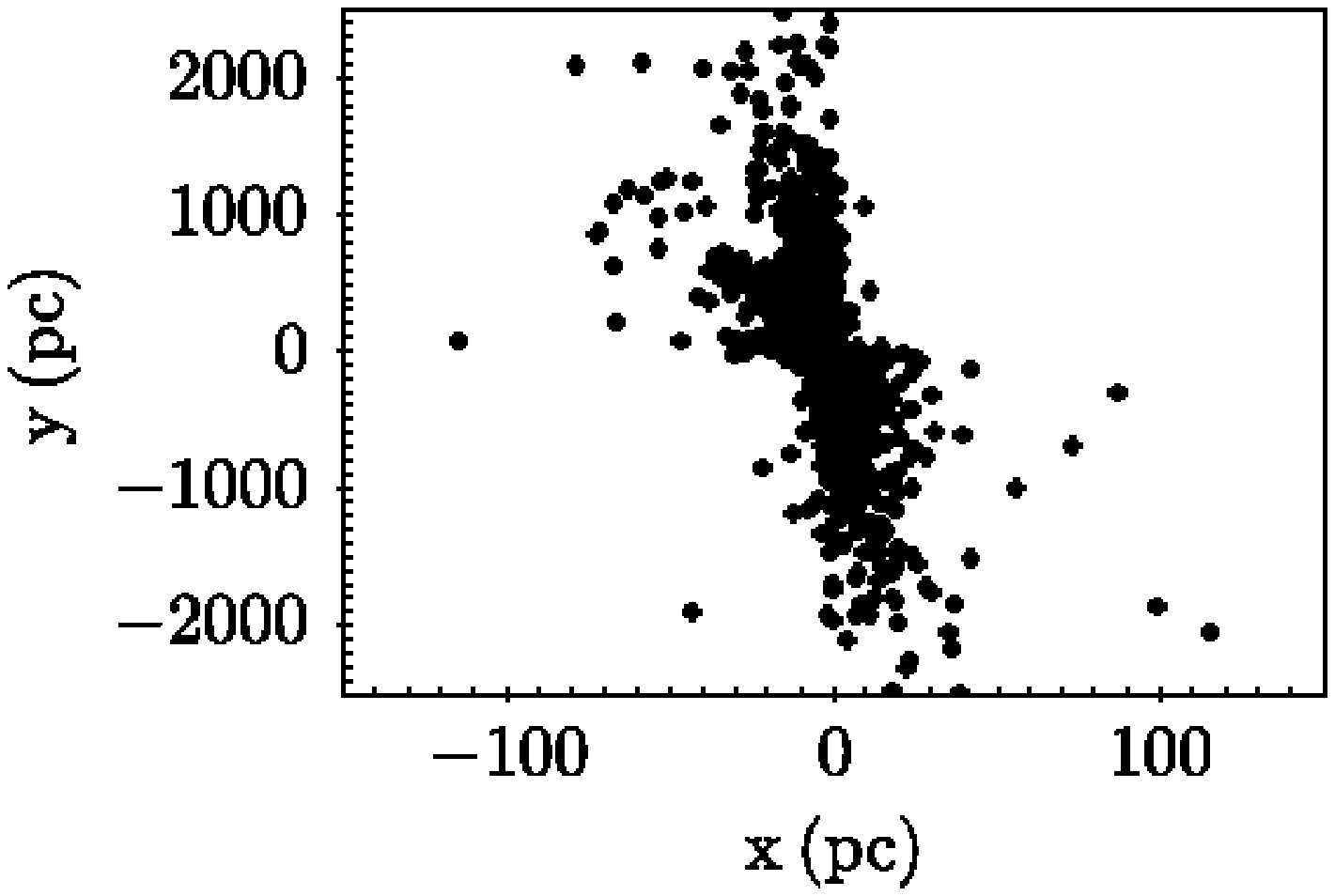}}
\subfigure{\includegraphics[width=0.48\columnwidth]{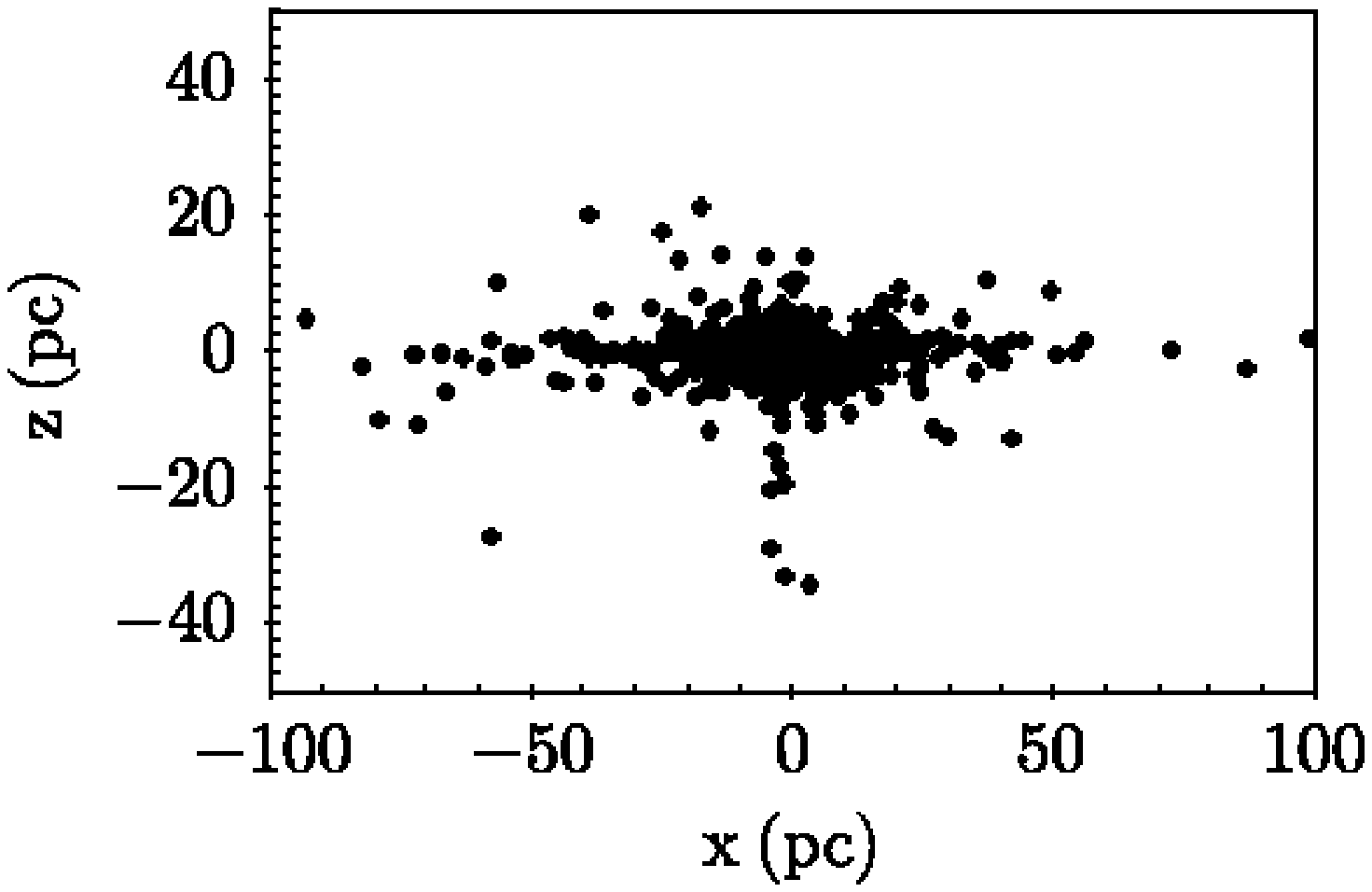}} \\ \subfigure{\includegraphics[width=0.48\columnwidth]{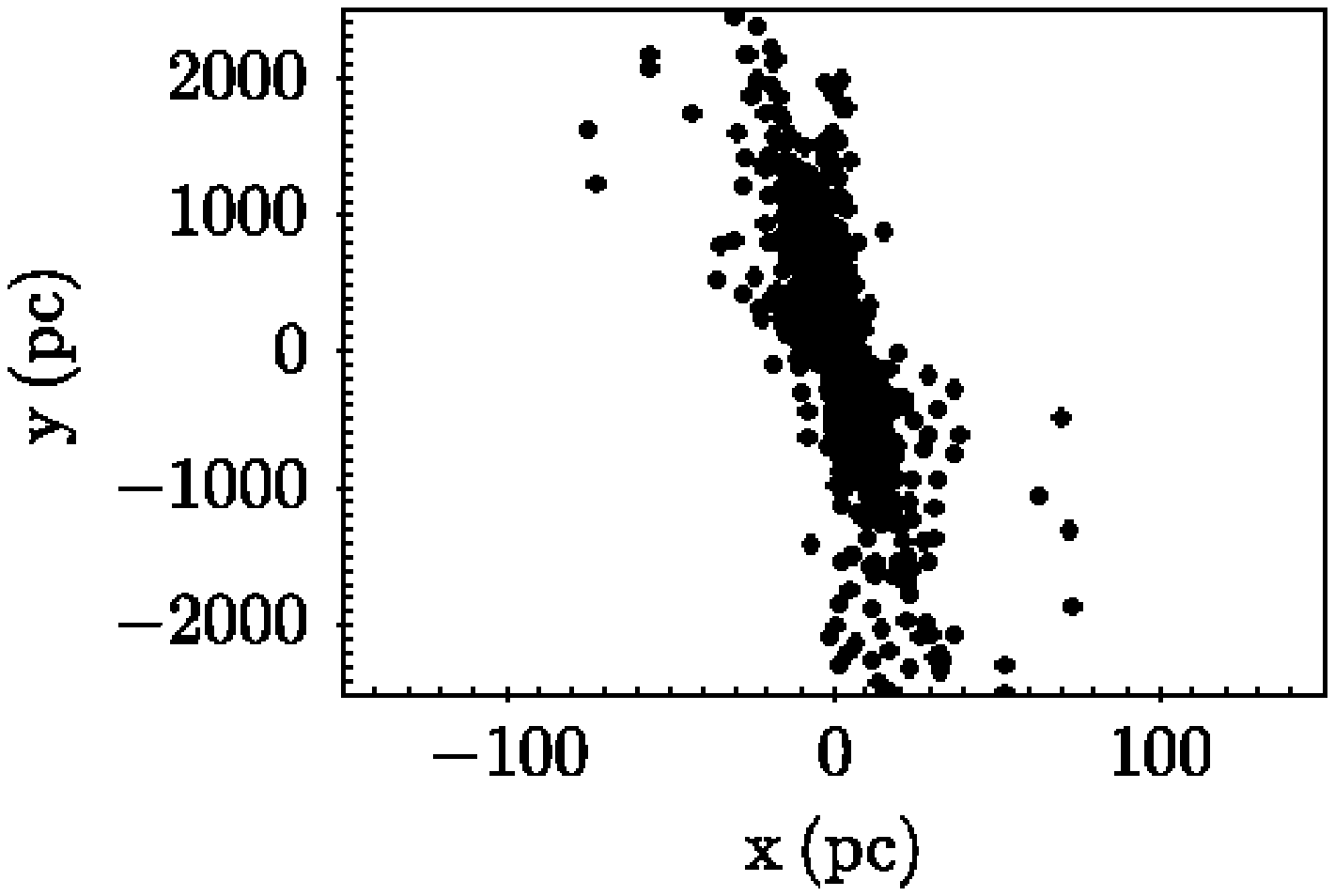}}
\subfigure{\includegraphics[width=0.48\columnwidth]{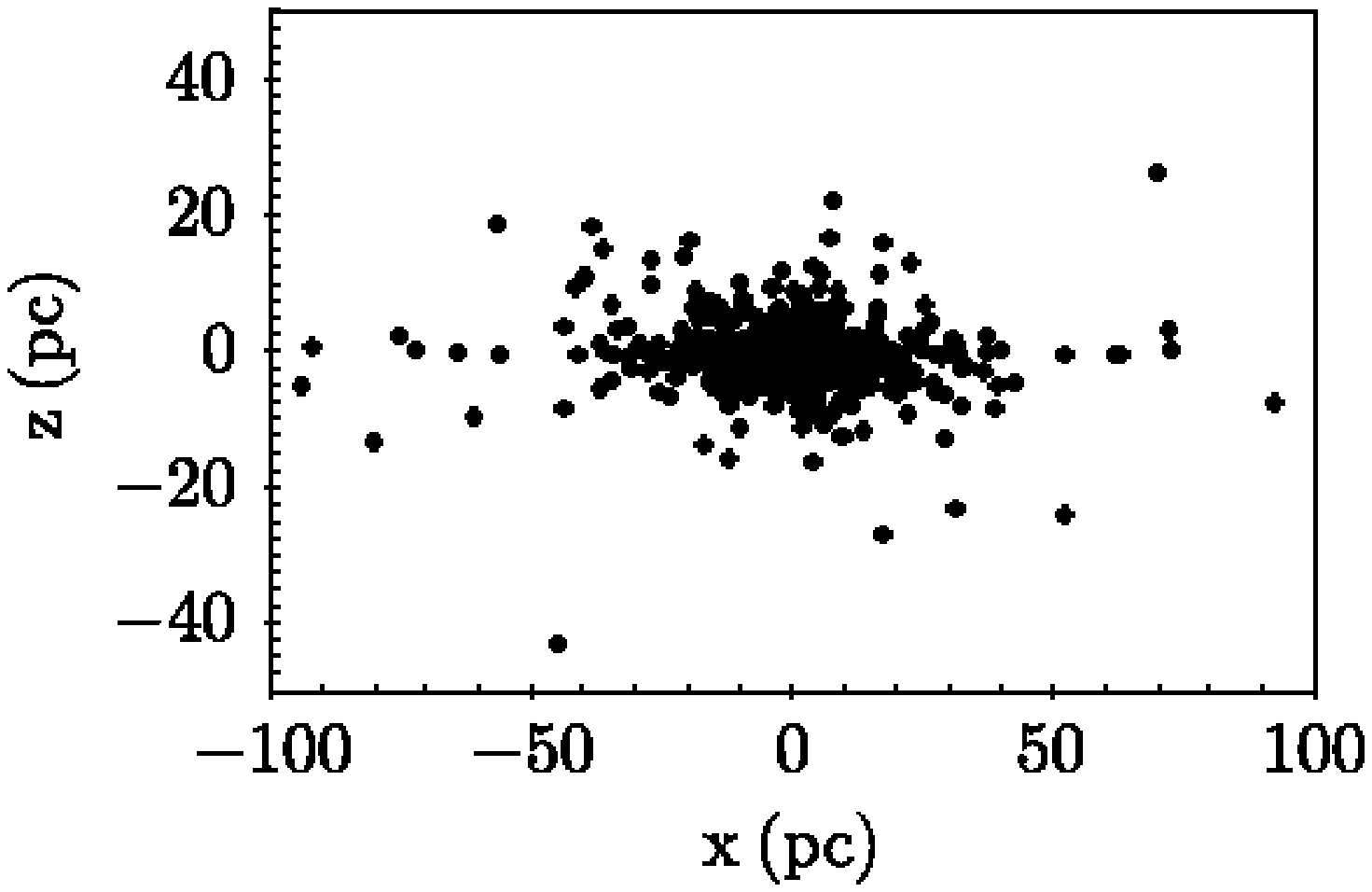}}
\caption{{\small Final distribution of the particles in the $x$-$y$ (left) and $x$-$z$ (right) planes, for the Samples 1 (top), 2 (middle) and 3 (bottom).}}
\label{figesc2}
\end{figure}

The surviving Oort Cloud has been analyzed in other works (e.g., Fouchard et al. 2017) and it is not the objective of our work. Instead, in this section, we made a theoretical analysis of the dynamic of unbound and ejected particles. Figure \ref{figesc2} shows the final distribution of the ejected particles in the Cartesian reference system, where we can see that such particles are moving around the Galaxy following the Sun with a small relative velocity. We can see a large structure which extends by 4 kpc (the half-path of the distance to the Galactic center) along the direction of Galactic rotation ($y-axis$) in a quasi-symmetric extension. 

However, the effect of the Galactic environment in our simulations is oversimplified for these large structures. We do not include perturbations from passing stars and molecular clouds, whose effect will lead to a dispersion of the particles in $x$-  and $z$-direction. Moreover, we have considered a non-migrating Sun in the solar neighborhood while a most realistic scenario, with a not fixed Sun and taking into account the influence of spiral arms, will be able to produce an important scattering of particles(Brasser et al. 2010;
Kaib et al. 2011; Martinez-Barbosa et al. 2016, 2017). So, this population of ejected particles must be more extended and maybe it could not exist at all.

\begin{figure}
% \begin{center}
%\epsfig{figure=maximo_evolucion_tot.eps,width=0.99\columnwidth}
\centering
\subfigure{\includegraphics[width=0.99\columnwidth]{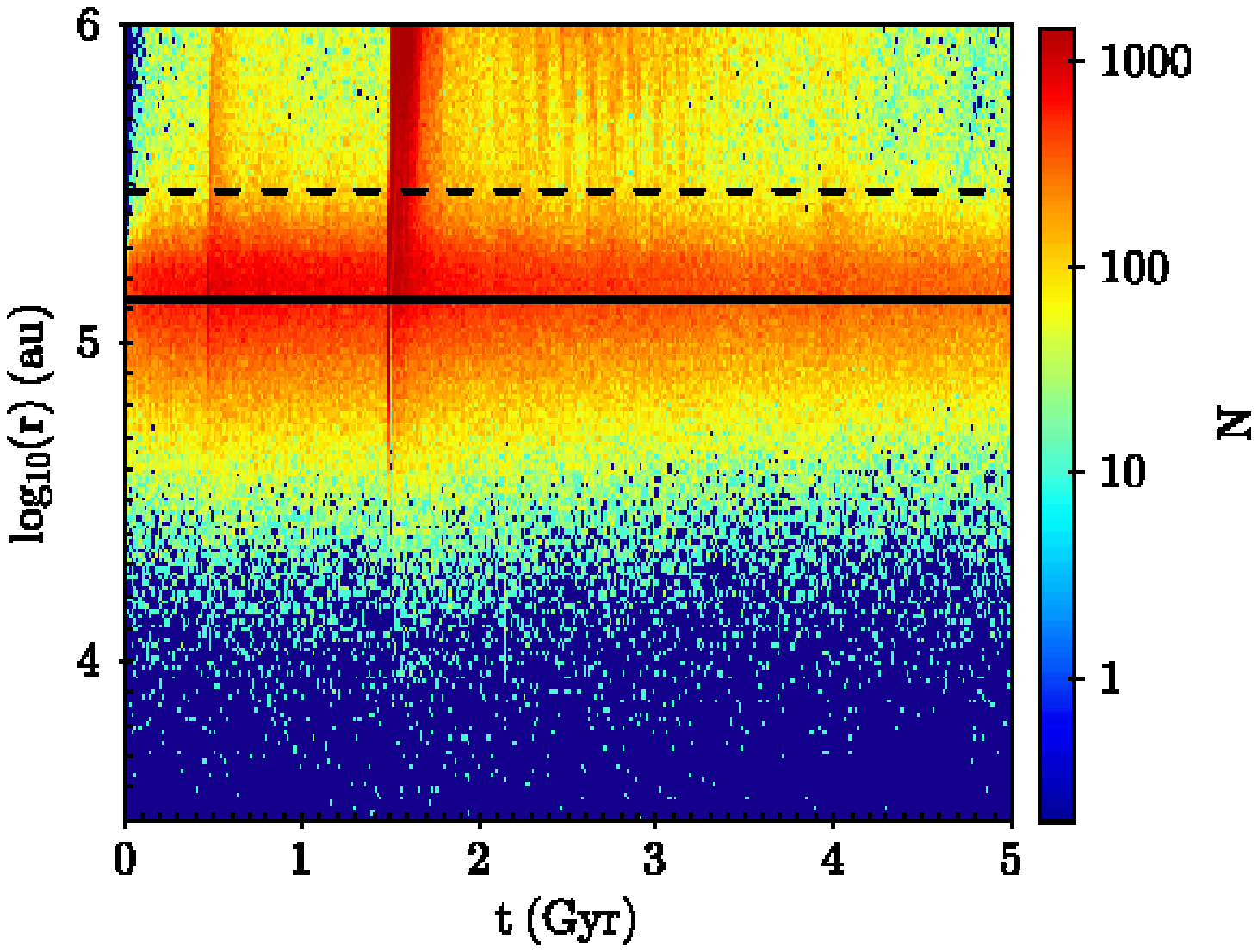}}
\subfigure{\includegraphics[width=0.99\columnwidth]{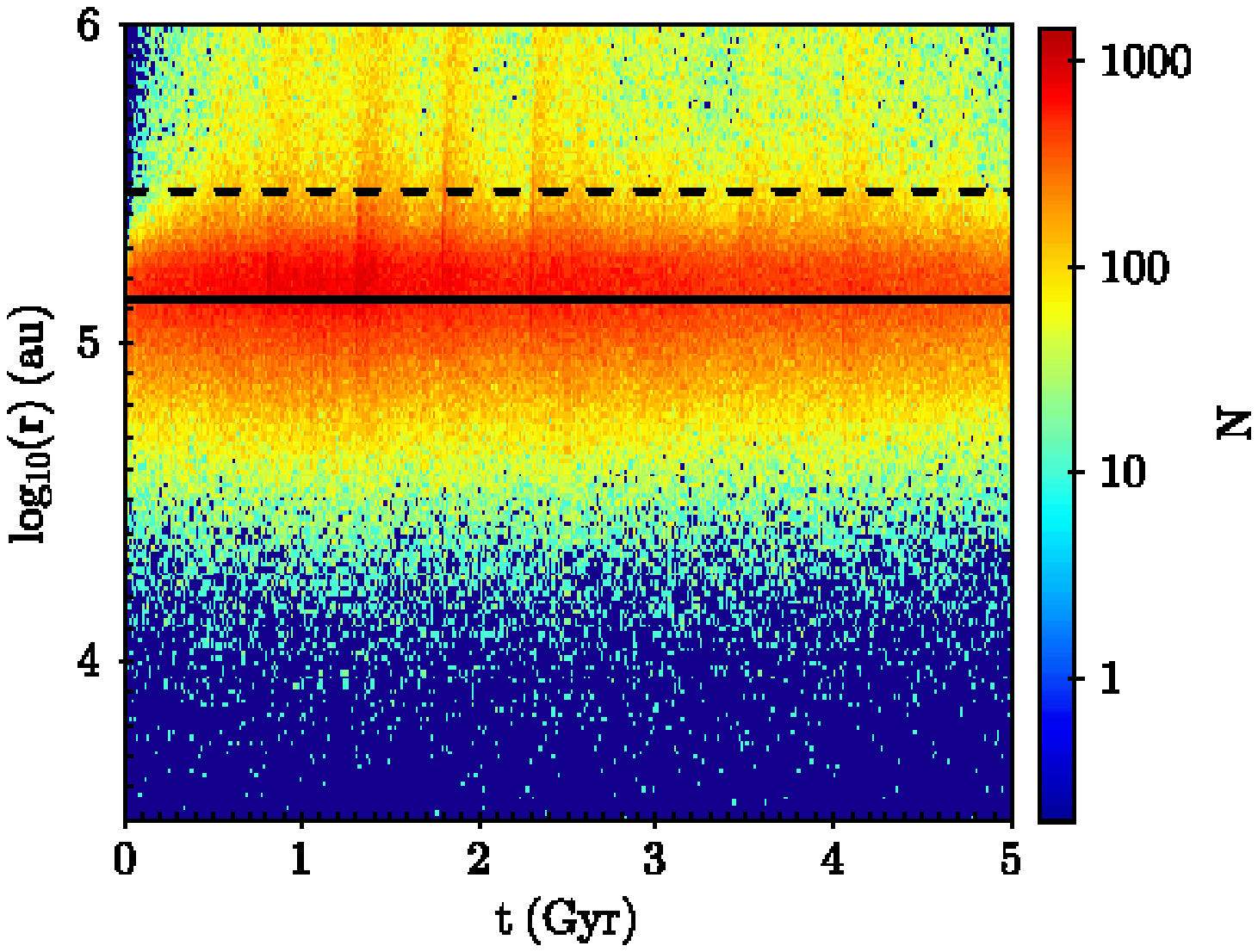}}
\subfigure{\includegraphics[width=0.99\columnwidth]{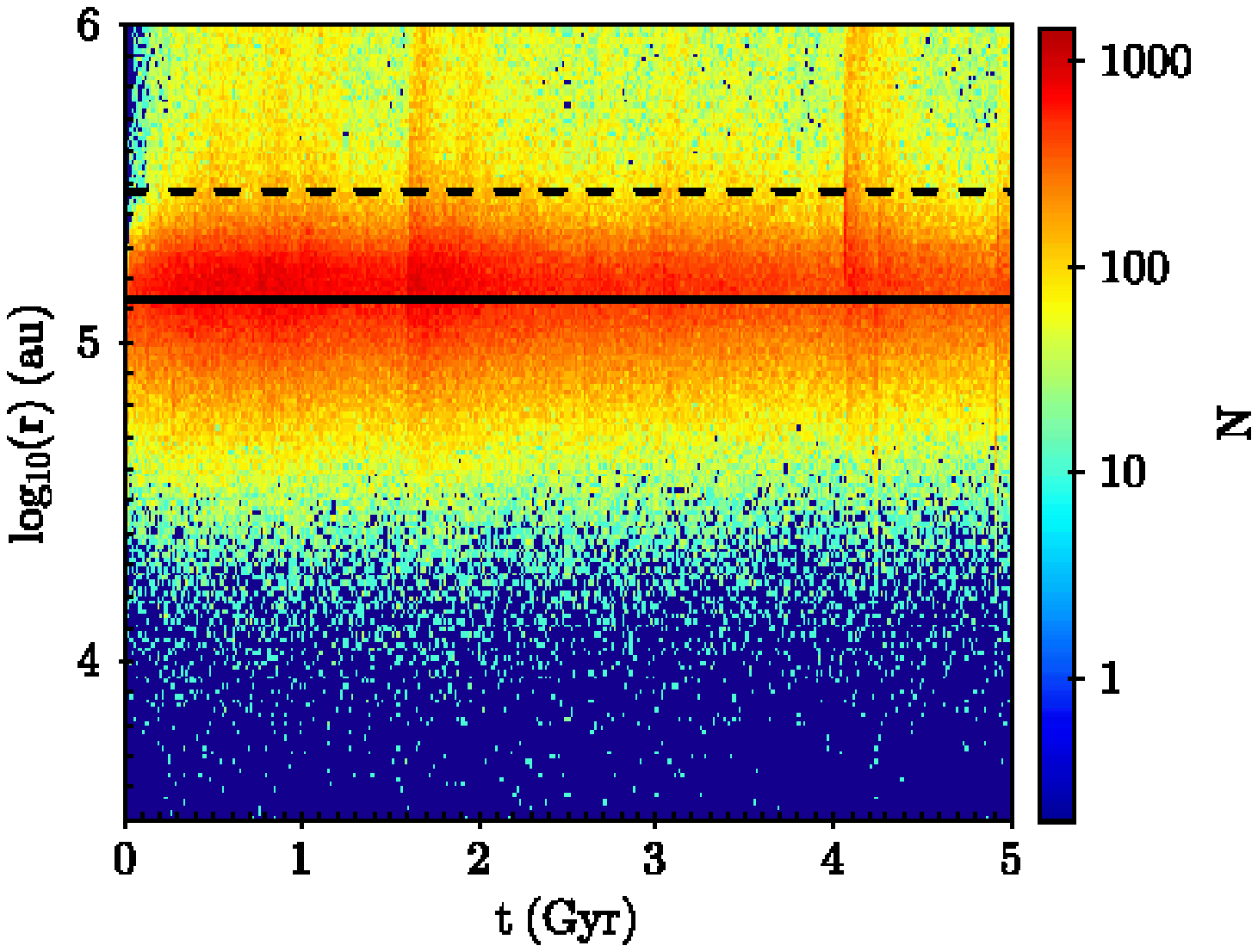}}
\caption{{\small Evolution of the distribution of the heliocentric distance $r$ of unbound particles ($C_J > C_{crit}$). The tidal radius, which divides regions $U_1$ and $U_2$, is shown with a dotted black line, while a half of the tidal radius is indicated with a continuous black line. From top to bottom, Samples 1, 2 and 3.}}
% \end{center}
\label{fig19}
\end{figure}

The other interesting group of objects is that of unbound but still not ejected particles. This group is conformed mainly by the particles in the region $U_1$ with distances between 10$^4$ au and $r_J$, see the black histograms in Fig. \ref{fig3}. Fig. \ref{fig19} shows the evolution of particles with $C_J > C_{crit}$ in the range between 10$^4$ and 10$^6$ au. We can see a constant escape of particles, but close to a half of the Jacobi radius is interesting to find a high density of particles. This region is permanently replenished by the successive stellar passages, and even in the case of a more disturbing sequence as Sample 1, the region is staying present.  This result is very interesting because indicates a dynamical structure around the Sun, where the flux of particles being ejected define a non-permanent population of unbound particles, which we call in this article: the \textit{unbound Oort cloud} (UOC, hereafter). 

The presence of the dynamical structure of the UOC is the consequence of the dynamical times involved in the problem. In absence of stellar passages, the particles evolve under the influence of the Keplerian and Galactic potentials. In this case, the unbound particles at $r \sim r_J$ can evolve in two directions: i) towards the Solar System with a Keplerian period of about 10 Myr, or ii) toward the interstellar space with a secular evolution governed by the Oort constant $A_G$ (see eq. \ref{eq8}), which indicates a constant secular increase of $r$ with a rate proportional to 0.01 Myr$^{-1}$. However, the average time between passages with our frequency of encounters is $\sim$ 0.1 Myr, or $\sim$ 0.05 Myr with the rate of encounters of the Gaia DR2 data  (Bailer-Jones et al. 2018). This difference of at least two orders of magnitude between the times of the two effects involves implies that the UOC can be considered as a frozen cloud of particles, which is continuously disturbed by the kick of stellar passages. The same is true for the most external particles of the Oort Cloud with $C_J \sim C_{crit}$. Then, every stellar passage moves some particles from the Oort Cloud toward the UOC, and at the same time, it moves some particles of the UOC towards the interstellar space. In this way, the UOC win and loose particles with every stellar passage, with a similar rate of gain and loss according to our results. In the next section we show that the UOC defines a natural boundary for the Oort Cloud, which is consistent with the escape radius formula for wide binary stars given by  Feng \& Jones (2017).

\subsection{The  outer limit of the  Oort Cloud}

In this section, we perform a dynamical analysis to find an external limit for the Oort Cloud. In the outer Oort Cloud the Galactic tide become important and we can not use the keplerian criteria of the eccentricity to define an unbound particle. So, following the criteria of Jiang \& Tremaine (2010) for a bound object (i.e., $C_J < C_{crit}$ and $r < r_J$), we can define that a massless particle is said to be stable only when it remains bounded to the Sun until the end of the simulation. Then, we define as unstable the unbound particles that remain close to the Sun, in the region $U_1$, because eventually they will increase their heliocentric distance (see Sect. \ref{evol}).

We perform a numerically study about the stability of particles disturbed by the Galaxy around the Sun, following the numerical studies about the stability of orbits in a restricted three-body problem (e.g., Wiegert \& Holman 1997; Holman \& Wiegert 1999; Ramos et al. 2015), but these results indicate that are necessary at least 300 binary periods to define the stability limit (e.g., Rabl \& Dvroak 1988; Holman \& Wiegert 1999; Ramos et al. 2015;
Calandra et al. 2018). Instead, for a period similar to the age of the Solar System the Sun has completed less than 30 revolutions around the Galaxy, which is less than a 10 \% of the necessary time to define a limit. Therefore, the limit that we could find is not stable and it will be reduced in the future, so we can only define a temporary limit for 5 Gyr.

\begin{figure}
\centering
\subfigure{\includegraphics[width=0.99\columnwidth]{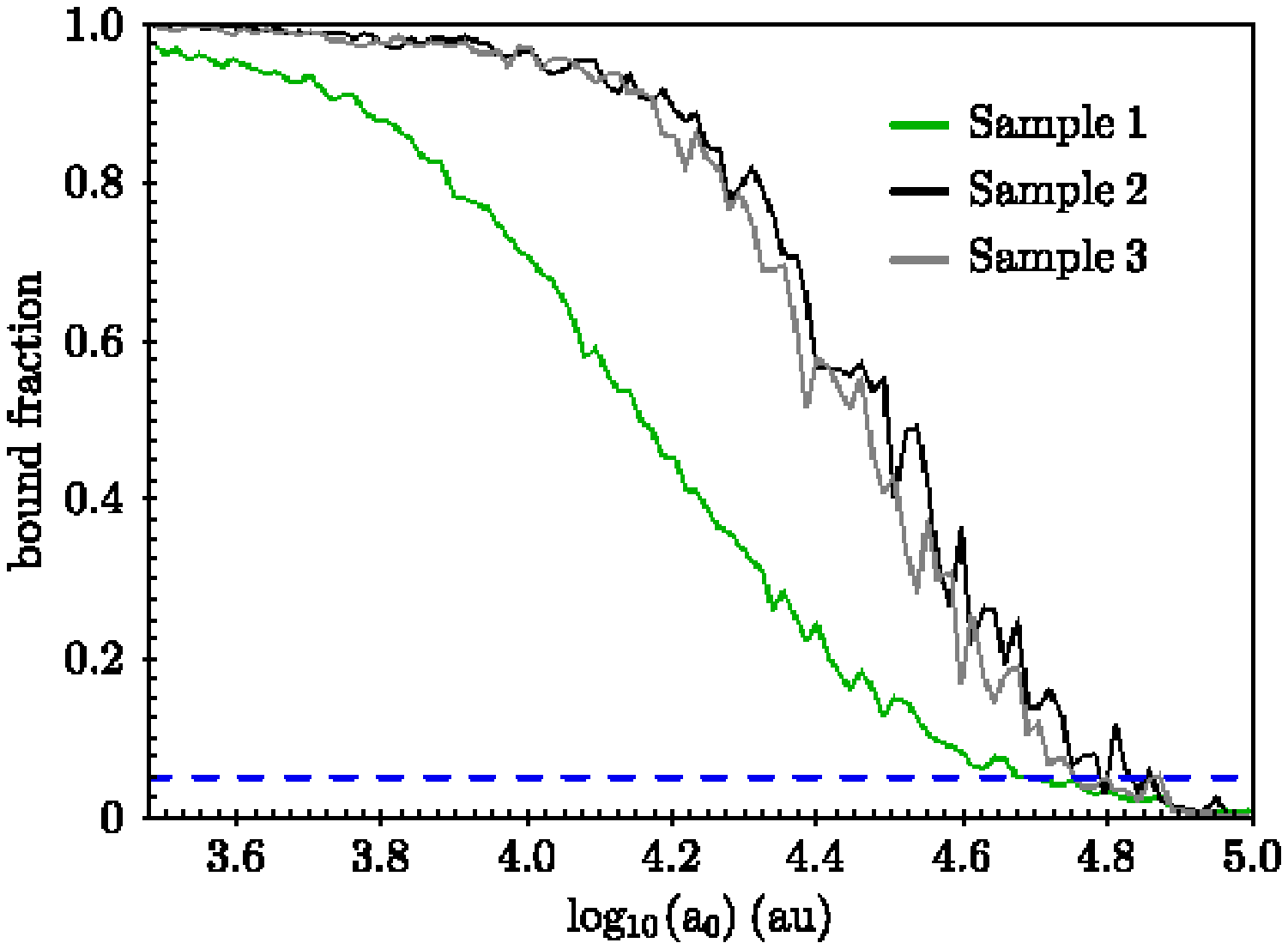}}
\subfigure{\includegraphics[width=0.99\columnwidth]{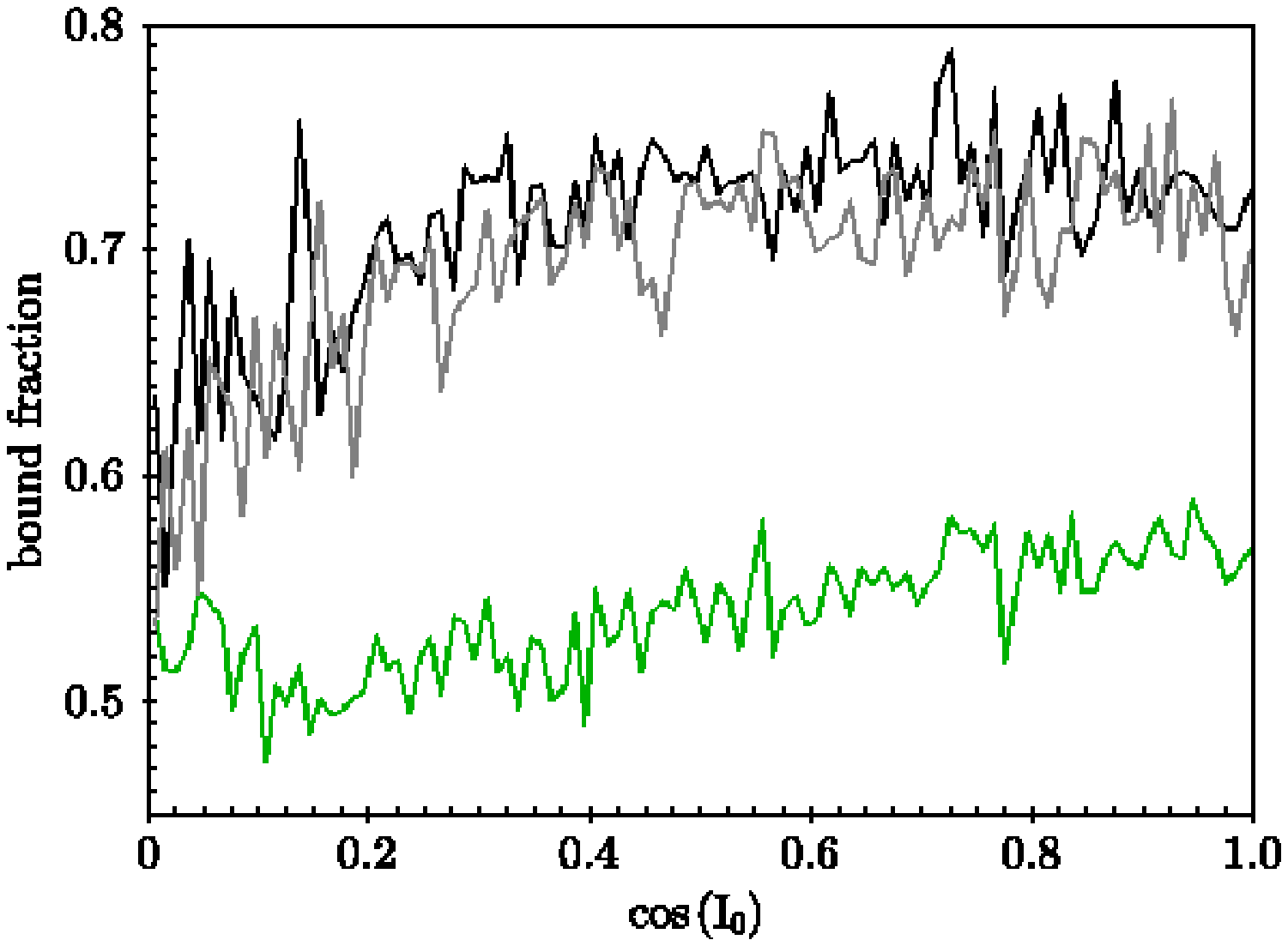}}
%\subfigure{\includegraphics[width=0.99\columnwidth]{rmenor_c0.eps}}
\caption{{\small  Fraction of bound particles (i.e., $C_J < C_{crit}$) with $r < r_J$ as function of the initial semimajor axis ($a_0$, top panel), and cosine of initial inclination ($\cos I_0$, bottom panel). Samples 1, 2 and 3 are indicated by green, black and grey lines, respectively. Blue dashed lines in top panel correspond to the fraction 0.05.  }}
\label{figlim2}
\end{figure}

Fig. \ref{figlim2} shows the fraction of bound particles that survive all the integration time, as function of i) the initial semimajor axis ($a_0$, top panel), and ii) the cosine of the initial inclination about the Galactic plane ($\cos I_0$, bottom panel). We have not include an analysis in function of the initial eccentricity because we do not find a dependence. The three samples are indicated by green (Sample 1), black (Sample 2) and grey (Sample 3) lines.

Our results indicate that the stability of the particles has a strong dependence with the initial heliocentric distance, and a slight dependence with the initial orientation of the orbit. Both quantities are dynamically important because their relation with the Jacobi constant regulates the stability of the particles (see eq. \ref{eq4}). So, from the set of particles with an initial value of Jacobi constant close to $C_{crit}$, those with larger initial inclinations can be ejected more easily due to the last term of the effective potential (eq. \ref{eq4}). Therefore, the Galactic potential reduces the bond energy of particles with large values of $I_0$, which is in agreement with the theoretical model (eq. \ref{eq3}). Finally, from our results we do not find a dependence with the initial eccentricity of the particles, which seems to indicate that for the stability of the external Oort Cloud the orientation of the orbits is more important than its form.

However, although we find a dependence with the initial configuration, the Sample 1 shows a different dependence, which indicate that the stability of the particles is defined by the set of stellar passages and specially by the possibility of a "special event". The effectiveness of this mechanism to eject particles is seen in Fig. \ref{figlim2} where at the initial border of the external Oort Cloud (2 $\times$ 10$^4$ au) both Samples 2 and 3 have a 80 \% of bound particles, which quickly decrease to 50 \% at 3 $\times$ 10$^4$ au. Instead, Sample 1 has 40 \% and 15 \% of bound particles at 2 $\times$ 10$^4$ au and 3 $\times$ 10$^4$ au, respectively. Moreover, the dependence with the inclination is most evident for Samples 2 and 3, while the Sample 1 is almost independent of the orientation of the orbit. These differences indicate that when the disturbing effect of the stellar passages increase (e.g., Sample 1), the Oort Cloud is more affected, and hence the dependence with the initial inclination of the survival particles decrease. 

Therefore, our results shown that the stability or the survival time of particles is defined by a interaction between the Galactic potential and the stellar passages in a synergy effect of ejection similar to that described by Rickman et al. (2008) for the injection of objects. The stellar passages are the perturbative effect, and the increase in the number of passages will increase the quantity of particles that can be ejected. However, to define when a particle is unbound from the Sun we need to consider the Galactic potential. In the model considered in this work  the particle is unbound when its Jacobi constant is larger than the critical value (see Sect. \ref{model}). Therefore, the stellar passages have to increase the value of $C_J$ to be able to eject the particles.  

Moreover, it is important to define how the stability limit is calculated. As our results shown, there is not a defined limit between bound and unbound orbits, but there is a \textit{transition region} that separates the stable from the unstable domains, like in the restricted three-body problem (Ramos et al. 2015). Then, we can define as "stability limit" the initial semimajor axis in the transition region where the bound fraction is 0.5, and considering the three samples we can calculate an average value. So, we obtain a mean stability limit for the transition region in $\sim$ 2 $\times$ 10$^4$ au ($log_{10}\,(a_0) \sim$ 4.3), which is in agreement with the limit that we can derive from previous works. For example, for the new encounter rate of 19.7 Myr$^{-1}$ derived from Gaia DR2 data, the analitical formula of Feng \& Jones (2017) estimate the escape radius or stability limit in $log_{10}\,(a_0) \sim 4.2$.

On the other hand, we are interested in finding an outer limit for the extension of the Cloud. In this case, we have to considered a region where the percentage of particles is very low. It is worth to note that the fraction of bound particles beyond 6.5 $\times$ 10$^4$ au is less than 0.05 and it decrease to 0 at 10$^5$ au. Moreover, unlike the stability limit, the unbound fraction is less than 0.05 at the same distance for the three Samples. Then, regardless the occurrence of a "special event" during the age of the Solar System the percentage of survival particles is less than 5 \% for the objects with initial semimajor axis larger than 6.5 $\times$ 10$^4$ au. Therefore, the stability for those particles is very low and we can qualitatively define $a_{crit} \sim$ 6.5 $\times$ 10$^4$ au as the outer limit for the Oort Cloud. Although, this boundary may be scaled if we take into account the approximations assumed in our simulations (see Sect. \ref{model}).

Finally, we show the importance of the Galactic potential for the stability of the particles using as illustrative example some real objects. From the JPL database of the NASA (http://ssd.jpl.nasa.gov) we found 14 comets with $a>$ 2 $\times$ 10$^4$ au, which correspond to objects in the outer Oort Cloud. Table \ref{table2} shows the name, perihelion distance ($d$) and approximated semimajor axis of each one. Even though these are quasi-parabolic comets (i.e., $e >$ 0.999), they have $e \neq$ 1 and in the context of the Keplerian problem we can not assure that they are unbound from the Sun and that they will be ejected from the Solar System, but using the proposed outer limit, which consider the Galactic potential, we can identify in Table \ref{table2} those comets that are in fact unbound from the Sun, and will be ejected.

\begin{table}
\centering
\begin{tabular}{c | c | c   }
  Comet  name  &   $d$ (au)   & $a$ ($\times$ 10$^4$ au)     \\
          \hline 
           \hline 
              &   & \\
C/2014 R3 (PANSTARRS)             &   7.27    & 2.26 \\
C/1958 D1 (Burnham)                 &     1.32    & 2.32 \\
C/2017 T2 (PANSTARRS)              &     1.61   & 2.35 \\
C/1910 A1 (Great January comet)     &     0.12    & 2.58 \\
C/2002 J4 (NEAT)                 &   3.63    & 2.89 \\
C/2001 C1 (LINEAR)                 &     5.10    & 3.81 \\ 
C/1972 X1 (Araya)                 &   4.86    & 5.40 \\
C/1937 N1 (Finsler)                 &     0.86    & 5.75 \\
 $- - - - - - - - - - - - - -  - - -  $ & $-  - - -$ & $-  - - - - $ \\
C/2007 N3 (Lulin)                  &     1.21    & 7.24 \\
C/1992 J1 (Spacewatch)           &     3.00    & 7.71 \\
C/2008 C1 (Chen-Gao)             &     1.26    & 10.16 \\
C/2012 CH17 (MOSS)                 &     1.29   & 14.00 \\
C/2012 S4 (PANSTARRS)             &     4.34    & 25.22 \\
C/2015 O1 (PANSTARRS)             &   3.73    & 44.65 \\
%\hline
\end{tabular}
\caption{Orbital parameters of the known comets with $a>$ 2 $\times$ 10$^4$ au, which are been taken from the JPL database of the NASA (http://ssd.jpl.nasa.gov). Second and third columns indicate the perihelion distance and the semimajor axis, respectively. The dashed line separate those objects with $a>$ 6.5 $\times$ 10$^4$ au, which are unbound according to our limit ($a_{crit}$).}
\label{table2}
\end{table}

%full_name    e    q    a (au)

%     C/2014 R3 (PANSTARRS)                0,99967842    7,275480240     22624,7833392511
%     C/1958 D1 (Burnham)                0,999943    1,322689     23205,0701754041
%     C/2017 T2 (PANSTARRS)             0,99993136    1,613784708     23511,7837649219
%     C/1910 A1 (Great January comet)    0,999995    0,128975     25794,999999831
%     C/2002 J4 (NEAT)                    0,99987425    3,633722001     28898,3590800086
%     C/2001 C1 (LINEAR)                0,999866091    5,104317173     38117,8932833247
%     C/1972 X1 (Araya)                    0,99991        4,860748     54008,3111110904
%     C/1937 N1 (Finsler)                0,999985    0,862744     57516,2666662899
%     C/2007 N3 (Lulin)                 0,999983259    1,21225836     72415,0000708731
%     C/1992 J1 (Spacewatch)            0,999961    3,007006     77102,7179486957
%     C/2008 C1 (Chen-Gao)                0,999987578    1,26234336    101627,542418511
%     C/2012 CH17 (MOSS)                0,999990736    1,296092176    139913,507951386
%     C/2012 S4 (PANSTARRS)                0,999982758    4,348730017    252223,798332182
%     C/2015 O1 (PANSTARRS)                0,999991646    3,729683715    446485,036463333

%###########################################################################################

% graficos: 18pt serif latex

\section{Discussion and Conclusions}\label{conclu}

In this paper we developed a dynamical study about the stability of the objects of the outer Oort Cloud. We considered the particles of the Oort Cloud in three different samples, each one was affected by a different sequence of stellar passages during 5 Gyr. We also took into account the perturbative effect of the Milky Way's tidal field.

From our results, we find that in absence of a "special event" there are a percentage of the initial material which is lost by ejection and a similar percentage lost by injection to the planetary region. However, as we do not include planets, we can not make predictions for the injected material. Thus a failure to include planetary perturbations may bias the results of the Oort Cloud evolution as well as the ejected comet population, and we can only say that the Cloud has lost at least $\sim$ 17 \% of the initial objects during the Sun lifetime. Moreover, because of the linear increase in the rate of ejection of particles, we can predict that for the end of its life the Sun will lost at least one third of the initial Cloud.

Another important result  is the final distribution of the heliocentric distances of the particles of our Samples. We found a minimum in the density at a few times $r_J$, independent of the sequence of stellar passages. Interior to this minimum there is a peak in the density due to the initial distribution of the particles in the Cloud, and exterior to it there is another peak due to particles that are slowly drifting away. This result is similar to that found for wide binary stars, which allow us to arrive to the conclusion that perturbations from the Galactic tide in a binary system is independent of the mass of the pair, with similar results for a two-body problem and a restricted two-body problem.

We have found a mechanism for the ejection of particles which we called "continuous process", which allow us to conclude that the leak of particles from the Oort Cloud is by a consequence of a synergy effect of ejection due to the combined effects of the successive stellar passages and the Galactic potential similar to the observed for the synergy effect of injection. We also found that the most important orbital characteristic of the particles of the Cloud that regulates the ejection is the initial semimajor axis. Moreover, the simulations shown that the usual treatment of ejection of particles from the Oort Cloud is oversimplified. The particles are not ejected isotropically when  $C_J > C_{crit}$ and the separation exceeds the Jacobi radius $r_J$; instead they have a defined distribution modulated by the Galactic potential. Although, as the heliocentric distance increases ($r \gg r_J$) the Galactic perturbations disperse the distribution of the ejected particles.

Therefore, we can define the Oort Cloud as a dynamically complex region where a population of bound particles live together with a group of unbound particles, which we identify as the unbound Oort Cloud. The UOC is not a population, instead it is a group of particles in a dynamically unstable region, which is permanently replenished by the permanent flow of stellar passages. Then, all the Oort Cloud is a transition region where the fraction of bound particles falls from 1 to 0 as the heliocentric distance increases, and allows to define the stability limit as the distance from which the unbound particles become dominant. We have found that the heliocentric distance with a similar quantity of bound and unbound particles is $\sim$ 2 $\times$ 10$^4$ au ($log_{10}\,(a_0) \sim 4.3$), which is in agreement with the estimated limit in previous works. It is worth to note that this result is in agreement with the limit between inner and outer Oort Cloud, and allow us to define the external Cloud as an unstable region. 

Moreover, we have defined $a_{crit} \sim $ 6.5 $\times$ 10$^4$ au ($log_{10}\,(a_{crit}) \sim 4.81$) as the outer or external limit for the outer Oort Cloud, because beyond this limit the probability of survival for the particles is very low ($<$ 5 \%) for the three sequences of stellar passages considered in this work. However, this boundary may be scaled because the approximations assumed in our simulations for example the frequency of encounters and a non-migrating Sun (see Sect. \ref{model}).

% Finally, an application of our external limit indicate that almost half of the population of known comets coming from the outer Oort Cloud belong to the unbound Oort Cloud and they will be ejected from the Solar System.

Finally, the results of our simulations could be improved in several ways. Our simulations consider an analytical model for the Galactic tide, and a more realistic treatment of the potential and a different Galactic environment could affect the definition of a unbound particle. Moreover, we  do not include perturbations from planets  and passing molecular clouds , and our rate of encounter for stellar passages is less than that considered in other works, which could affect the position of the limits found in this work. Even so, we think that our most important contributions are not the exact position of a region or a limit, but instead we have been able to determine the existence of a transition region and the UOC, which define a phase transition between the Solar System and the interstellar space. Both dynamic structures are independent of our approximations, because there will always be a fraction of unbound particles that will not be ejected in the time between one stellar passage and the next, and the existence of such transitory population makes impossible to define an exact stability limit. The main modulator of these structures and the external limit of the Oort Cloud is the frequency of the stellar passages.

\section{REFERENCES}

Bacci P., et al. 2017, MPEC Circ.. MPEC 2017-U181

Bailer-Jones C. A. L., Rybizki J., Andrae R., Fouesneau M., 2018, A\&A,616, 37

Binney J., Tremaine S., 2008, Galactic Dynamics: Second Edition. Princeton University Press

Brasser R., Morbidelli A., 2013, Icarus, 225, 40

Brasser R., Duncan M. J., Levison H. F., 2006, Icarus, 184, 59

Brasser R., Higuchi A., Kaib N., 2010, A\&A, 516, 72

Brunini A., Fernandez J., 1996, A\&A, 308, 988

Byl J., 1983, Earth, Moon and the Planets, 29, 121

Calandra M. F., Correa-Otto J., Gil-Hutton R., 2018, A\&A, 611, 73

Correa-Otto J., Gil-Hutton R., 2017, A\&A, 608, 116

Correa-Otto J., Calandra M. F., Gil-Hutton R., 2017, A\&A, 600, 59

Dones L., Weissman P. R., Levison H. F., Duncan M. J., 2004, in Johnstone
D., Adams F. C., Lin D. N. C., Neufeeld D. A., Ostriker E. C., eds, Astronomical Society of the Pacific Conference Series Vol. 323, Star Formation in the Interstellar Medium: In Honor of David Hollenbach. p. 371

Dones L., Brasser R., Kaib N., Rickman H., 2015, Space Sci. Rev., 197, 191

Duncan M., Quinn T., Tremaine S., 1987, AJ, 94, 1330

Dybczynski P. A., 2002, A\&A, 396, 283

Feng F., Bailer-Jones C. A. L., 2014, MNRAS, 442, 3653

Feng F., Bailer-Jones C. A. L., 2015, MNRAS, 454, 3267

Feng F., Jones H. R. A., 2017, MNRAS, 474, 4412

Fernandez J. A., 1980, Icarus, 42, 406

Fernandez J. A., Brunini A., 2000, Icarus, 145, 580

Fouchard M., Froeschlé C., Rickman H., Valsecchi G. B., 2011a, Icarus, 214, 334

Fouchard M., Rickman H., Froeschle C., Valsecchi G. B., 2011b, A\&A, 535, 13

Fouchard M., Rickman H., Froeschle C., Valsecchi G. B., 2014, Icarus, 231, 110

Fouchard M., Rickman H., Froeschle C., Valsecchi G. B., 2017, Icarus, 292, 218

Francis P. J., 2005, ApJ, 635, 1348

Garcia-Sánchez J., Weissman P., Preston R., et al. 2001, A\&A, 379, 634

Hanse J., Jilkova L., Portegies Zwart S. F., Pelupessy F. I., 2018, MNRAS, 473, 5432

Heggie D. C., 2001, in Steves B. A., Maciejewski A. J., eds, The Restless Universe. pp 109–128 (arXiv:astro-ph/0011294)

Heisler J., Tremaine S., 1986, Icarus, 65, 13

Hills J. G., 1981, AJ, 86, 1730

Holman M. J., Wiegert P. A., 1999, AJ, 117, 621

Hut P., Tremaine S., 1985, AJ, 90, 1548

Jakubík M., Neslusan L., 2009, CASP, 39, 85

Jiang Y. F., Tremaine S., 2010, MNRAS, 401, 977–994

Kaib N., Quinn T., 2008, Icarus, 197, 221

Kaib N. A., Quinn T., 2009, Science, 325, 1234

Kaib N., Roskar R., Quinn T., 2011, Icarus, 215, 491

Martínez-Barbosa C. A., Brown A. G. A., Boekholt T., Portegies Zwart S. F., Antiche E., Antoja T., 2016, MNRAS, 457, 1062

Martinez-Barbosa C. A., Jikova L., Portegies Zwart S. F., Brown A. G. A., 2017, MNRAS, 464, 2290

Meech K., et al. 2017a, MPEC Circ.. MPEC 2017-U183

Meech K., Weryk R., Micheli M., et al. 2017b, Nature, 552, 378

Ninkovic S., Trajkovska V., 2006, Serb. Astron. J., 172, 17

Oort J. H., 1950, Bull. Astron. Inst. Netherlands, 11, 91

Rabl G., Dvroak R., 1988, A\&A, 191, 385

Ramos X. S., Correa-Otto J. A., Beauge C., 2015, CeMDA, 123, 453

Reid I. N., Gizis J. E., Hawley S. L., 2002, AJ, 124, 2721

Rickman H., 1976, Bulletin of the Astronomical Institutes of Czechoslovakia, 27, 92

Rickman H., Fouchard M., Froeschly C., Valsecchi G. B., 2008, CeMDA, 102, 111

Tsiganis K., Gomes R., Morbidelli A., Levison H. F., 2005, Nature, 435, 459

Vokrouhlicky D., Nesvorny D., Dones L., 2019, AJ, 157, 181–208

Weissman P. R., 1996, Completing the Inventory of the Solar System, Astronomical Society of the Pacific Conference Proceedings, vol. 107. Rettig T. W., Hahn J. M., eds., pp. 265-288

Wiegert P. A., Holman M. J., 1997, AJ, 113, 1445

\section*{Acknowledgements}

We thank Ricardo Gil-Hutton for numerous suggestions/correction on this paper. We thank the referee Fabo Feng for valuable suggestions/correction which help to improve the manuscript.
The authors gratefully acknowledges partial financial support by CONICET through PIP 112-201501-00525, and by CICITCA UNSJ, through the projects 21/E1079 (2018-2019) and PROJOVI (2018-2019).
% The authors are grateful to anonimus referee for numerous suggestions/correction on this paper.

%%%%%%%%%%%%%%%%%%%%%%%%%%%%%%%%%%%%%%%%%%%%%%%%%%

%%%%%%%%%%%%%%%%%%%% REFERENCES %%%%%%%%%%%%%%%%%%

% The best way to enter references is to use BibTeX:

%\bibliographystyle{mnras}
%\bibliography{nube} % if your bibtex file is called example.bib

% Alternatively you could enter them by hand, like this:
% This method is tedious and prone to error if you have lots of references
%\begin{thebibliography}{99}
%\bibitem[\protect\citeauthoryear{Author}{2012}]{Author2012}
%Author A.~N., 2013, Journal of Improbable Astronomy, 1, 1
%\bibitem[\protect\citeauthoryear{Others}{2013}]{Others2013}
%Others S., 2012, Journal of Interesting Stuff, 17, 198
%\end{thebibliography}

%%%%%%%%%%%%%%%%%%%%%%%%%%%%%%%%%%%%%%%%%%%%%%%%%%

%%%%%%%%%%%%%%%%% APPENDICES %%%%%%%%%%%%%%%%%%%%%

%\appendix

%\section{Some extra material}

%If you want to present additional material which would interrupt the flow of the main paper,
%it can be placed in an Appendix which appears after the list of references.

%%%%%%%%%%%%%%%%%%%%%%%%%%%%%%%%%%%%%%%%%%%%%%%%%%

% Don't change these lines
\bsp	% typesetting comment
\label{lastpage}
\end{document}